\documentclass[aps,physrev,twocolumn,superscriptaddress]{revtex4-2}
\usepackage{graphicx}%
\usepackage{amsmath}%
\usepackage{hyperref}%
\usepackage{xcolor}%
\usepackage{amssymb}

\begin{document}

%\preprint{}

\title{Adaptive-basis sample-based neural diagonalization for quantum many-body systems}


%\author{Simone Cantori$^{1,2}$, Luca Brodoloni$^{1,2}$, Edoardo Recchi$^{1}$, Emanuele Costa$^{3,4}$, Bruno Juli\'a-D\'iaz$^{3,4}$, Sebastiano Pilati$^{1,2}$}
%\email[]{simone.cantori@unicam.it}
%\affiliation{$^1$School of Science and Technology, Physics Division, University of Camerino, I-62032 Camerino (MC), Italy\\
%$^2$INFN-Sezione di Perugia, 06123 Perugia, Italy\\
%$^3$Departament de F\'isica Qu\`antica i Astrof\'isica, Facultat de F\'isica, Universitat de Barcelona, 08028 Barcelona, Spain\\
%$^4$Institut de Ci\`encies del Cosmos de la Universitat de Barcelona, ICCUB, 08028 Barcelona, Spain
%}

\author{Simone Cantori}
\email[]{simone.cantori@unicam.it}
\affiliation{School of Science and Technology, Physics Division, University of Camerino, I-62032 Camerino (MC), Italy}
\affiliation{INFN-Sezione di Perugia, 06123 Perugia, Italy}
\author{Luca Brodoloni}
\affiliation{School of Science and Technology, Physics Division, University of Camerino, I-62032 Camerino (MC), Italy}
\affiliation{INFN-Sezione di Perugia, 06123 Perugia, Italy}
\author{Edoardo Recchi}
\affiliation{School of Science and Technology, Physics Division, University of Camerino, I-62032 Camerino (MC), Italy}
\author{Emanuele Costa}
\affiliation{Departament de F\'isica Qu\`antica i Astrof\'isica, Facultat de F\'isica, Universitat de Barcelona, 08028 Barcelona, Spain}
\affiliation{Institut de Ci\`encies del Cosmos de la Universitat de Barcelona, ICCUB, 08028 Barcelona, Spain
}
\author{Bruno Juli\'a-D\'iaz}
\affiliation{Departament de F\'isica Qu\`antica i Astrof\'isica, Facultat de F\'isica, Universitat de Barcelona, 08028 Barcelona, Spain}
\affiliation{Institut de Ci\`encies del Cosmos de la Universitat de Barcelona, ICCUB, 08028 Barcelona, Spain
}
\author{Sebastiano Pilati}
\affiliation{School of Science and Technology, Physics Division, University of Camerino, I-62032 Camerino (MC), Italy}
\affiliation{INFN-Sezione di Perugia, 06123 Perugia, Italy}


%\date{\today}

\begin{abstract}
Accurately estimating ground-state energies of quantum many-body systems is still a challenging computational task because of the exponential growth of the Hilbert space with the system size. Sample-based diagonalization (SBD) methods address this problem by projecting the Hamiltonian onto a subspace spanned by a selected set of basis configurations. In this article, we introduce two neural network-enhanced approaches for SBD: sample-based neural diagonalization (SND) and adaptive-basis SND (AB-SND). Both employ autoregressive neural networks to efficiently sample relevant basis configurations, with AB-SND additionally optimizing a parameterized basis transformation so that the ground-state wave function becomes more concentrated. We consider different classes of basis transformations: single-spin and non-overlapping two-spin rotations, which are tractable on classical computers, and also more expressive global unitaries that can be implemented using quantum circuits. We demonstrate the effectiveness of these techniques on various quantum Ising models, showing that SND achieves high accuracy for concentrated ground states, while AB-SND consistently outperforms both SND and more conventional SBD methods, allowing entering regimes in which the ground state is not concentrated in the original computational basis. 
%In both methods, the accuracy improves as more configurations are sampled.
\end{abstract}

%\keywords{}

\maketitle

\section{Introduction}
The accurate calculation of ground-state properties of quantum many-body systems is one of the central challenges in quantum chemistry and condensed matter physics. The exponential growth of the Hilbert space with the system size makes exact solutions intractable for large systems, necessitating the development of approximate computational methods.
Deep learning methods have emerged as promising tools to address this challenge~\cite{RevModPhys.91.045002,Carrasquilla01012020,Kulik_2022}. For example, supervised learning approaches have been used to predict ground-state energies based on labeled training data~\cite{Dunjko_2018,PhysRevE.102.033301,C8SC04578J, doi:10.1021/acs.jctc.7b00577}. On the other hand, the introduction of neural quantum states (NQS), which represent wave functions using neural network (NN) architectures, has opened new possibilities for variational Monte Carlo simulations, circumventing the need of labeled data~\cite{Carleo_2017,PhysRevLett.124.020503,Barrett2022}.

In quantum chemistry, a standard approach to tackle the problem of the Hilbert space size is represented by selected configuration interaction methods~\cite{Abraham_2020,10.1063/5.0233542, doi:10.1139/cjc-2013-0017,PhysRevB.75.224503}. These employ predefined wave-function ansatzes, Monte Carlo sampling, or other empirical criteria to select a set of relevant basis configurations $| x^{(l)} \rangle$, labeled by the index $l$. 
%For these configurations, represented by bitstrings $x^{(l)}$ (i.e., elements of the computational basis, formally introduced in Sec.\ref{methods}), 
The corresponding Hamiltonian matrix elements $\langle x^{(l)} | H | x^{(m)} \rangle$ are evaluated to define a subspace Hamiltonian matrix. The ground-state energy is then approximated by computing the lowest eigenvalue of this matrix. 
Recently, these approaches have also been adopted in the context of quantum computing, under the name of Sample-based Diagonalization (SBD)~\cite{Robledo_Moreno_2025,Yoshioka_2025, kanno2023quantumselectedconfigurationinteractionclassical}. The key idea is to employ quantum circuits to sample relevant configurations, leading to what is dubbed Sample-based Quantum Diagonalization (SQD). In principle, quantum circuits might allow for the sampling of classically intractable distributions~\cite{arute2019quantum}.
Machine learning algorithms have also been used to select relevant configurations~\cite{Coe_2018,PhysRevB.111.035124,PhysRevLett.131.133002,Rano2023}.
Yet, the problem of how to efficiently truncate the Hilbert space, while minimizing the introduced approximation, is still open. In fact, SBD approaches are known to perform well only when the ground-state wave function is concentrated on the chosen computational basis~\cite{Robledo_Moreno_2025}, which means that its amplitudes are not negligible only on a small subset of basis elements. This strongly limits the regime of applicability of SBD methods.

%To emphasize the sampling perspective and its relevance to hybrid quantum-classical applications, we adopt the term Sample-based Diagonalization (SBD), inspired by recent developments in Sample-based Quantum Diagonalization (SQD)~\cite{Robledo_Moreno_2025,Yoshioka_2025}.  Recent advances have integrated deep learning to enhance the selection of configurations, thereby reducing the computational cost while maintaining the accuracy~\cite{Coe_2018,PhysRevB.111.035124,PhysRevLett.131.133002,Rano2023}. However, these SBD techniques become ineffective when the ground state is not sufficiently concentrated in the chosen basis, since they need to sample too many relevant configurations.

%This method achieves accuracies comparable to those obtained by sampling directly from the true ground state. 
%We further introduce a second method, called basis dependent - sample based neural diagonalization (AB-SND), which iteratively change the basis to make the ground state representation increasingly sparser. A schematic overview illustrating the functioning of both methods is presented in Fig.~\ref{fig1}.
%
\begin{figure*}
	\centering
	\includegraphics[width=\textwidth]{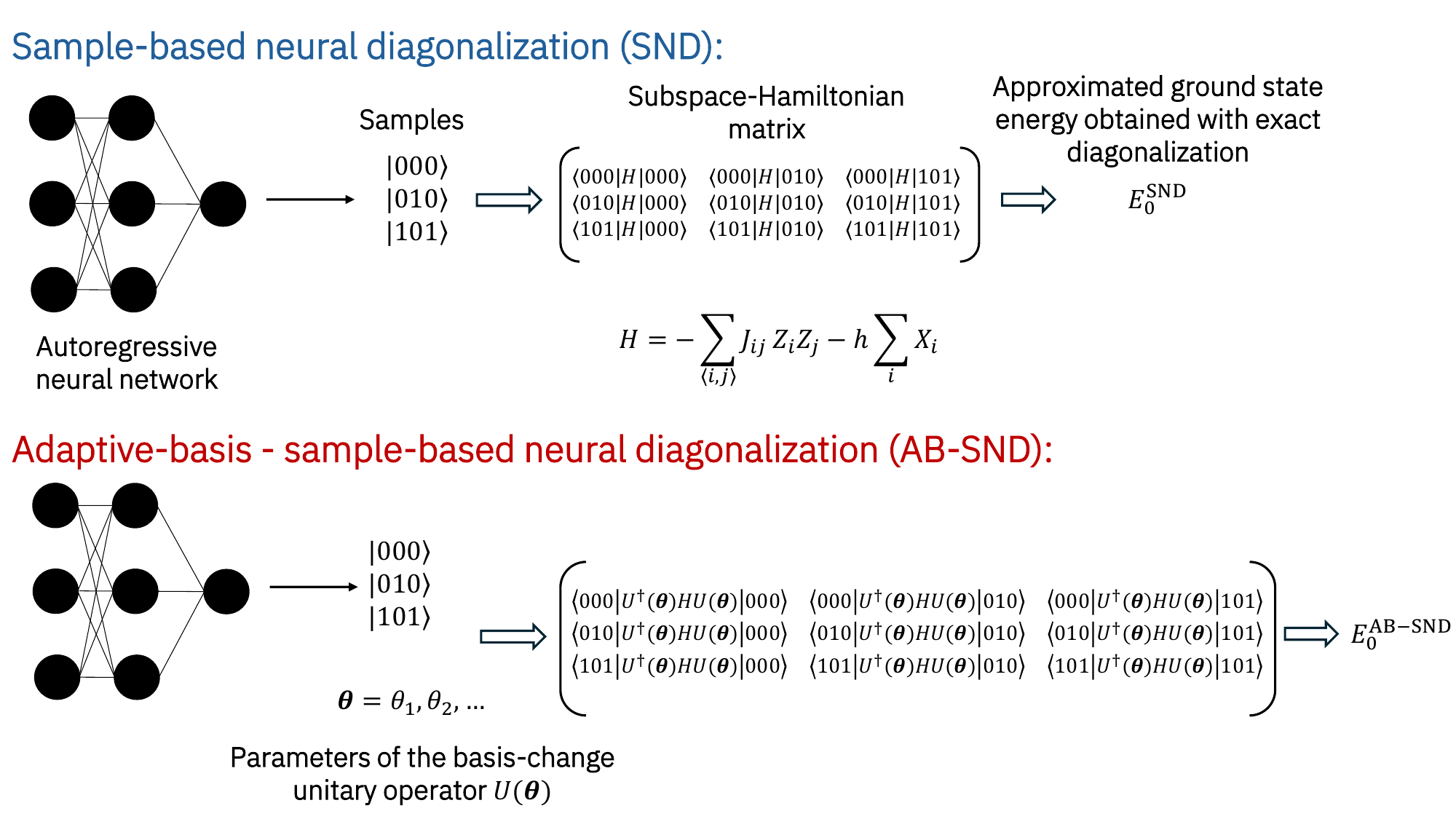}
	\caption{Scheme of the SND and AB-SND methods. In SND (top), an autoregressive neural network generates the bitstrings that define the subspace for diagonalization. In AB-SND (bottom), we also optimize some basis-transformation parameters $\boldsymbol{\theta}$, enabling us to perform sampling in a rotated basis where the ground state is more concentrated.} 
	\label{fig1}
\end{figure*}
In this article, we introduce two NN enhanced SBD approaches: sample-based neural diagonalization (SND) and its extension based on an adaptive basis, which we refer to as adaptive-basis SND (AB-SND). Both methods employ autoregressive NNs to efficiently sample basis configurations relevant for the estimation of the ground-state energy, as illustrated in Fig.~\ref{fig1}. While SND operates in a fixed computational basis, AB-SND incorporates a basis transformation, allowing for improved performance when the ground state is not concentrated in the original computational basis, a regime in which standard SBD techniques are doomed to fail. In our implementation, the basis change is performed using parameterized single-spin and two-spin rotations, which are efficiently computable on classical hardware. Additionally, we explore more expressive basis transformations, describing how quantum computers allow their implementation for large systems.
The testbeds we consider are one-dimensional (1D) and two-dimensional (2D) ferromagnetic quantum Ising models and a 2D quantum spin-glass model. These models are chosen because the concentration of their ground state can be tuned, allowing us to test SBD approaches in different regimes of computational hardness, and because the benchmark ground-state energy can be exactly computed.
Notably, we find that the AB-SND approach allows us to considerably extend the regime of applicability of SBD methods, in particular, when performed with the more general basis transformations.

The rest of this article is organized as follows: Sec.~\ref{methods} provides the details of the methodology of both SND and AB-SND. Sec.~\ref{results} presents our numerical results for different quantum Ising models, comparing the performance of the proposed approaches to standard SBD techniques based 
%on exact ground-state sampling or on 
on sampling from NQS variational ansatzes. In Sec.~\ref{qc}, we discuss the possibility of implementing AB-SND using quantum hardware, highlighting its potential for hybrid quantum-classical schemes. In Sec.~\ref{conclusion}, we summarize our findings and discuss future directions. Finally, the Appendices provide additional details on: the definition of the loss functions and their gradient calculations for SND (Appendix~\ref{app1}) and AB-SND (App.~\ref{app2}); the failure of standard SBD when the ground state is not concentrated (App.~\ref{app3}); the regime of maximum error in AB-SND (App.~\ref{app4}); the optimization of angle parameters via stochastic sampling (App.~\ref{app5}); the effective temperature scaling for efficient sampling of unique configurations (App.~\ref{app6}); the autoregressive NN architectures (App.~\ref{app7}).
%SND methods correspond to an autoregressive neural network (ANN) model employed as ansatz for SBD, as illustrated in Fig.~\ref{fig1}. 

\section{Methods}\label{methods}
As in standard SBD approaches, the SND methods we introduce hereafter aim to approximate the ground-state energy of a quantum system by projecting the Hamiltonian onto a subspace spanned by a selected set of basis configurations. In our implementation, these configurations are selected from the standard computational basis, which consists of tensor products of the single-qubit basis states $|0\rangle$ and $|1\rangle$, namely, the eigenstates of the Pauli-$Z$ operator. For a system of $N$ spins, the computational basis states $|x\rangle=|x_1x_2\dots x_N\rangle$ correspond to bitstrings $x$, where each component $x_i\in\{0,1\}$. Given a set of $S$ unique configurations, $\{|x^{(l)}\rangle\}_{l=1,\dots S}$, one constructs a subspace Hamiltonian by evaluating the matrix elements $\langle x^{(l)} | H | x^{(m)} \rangle$. The lowest eigenvalue $E$ of this subspace matrix provides a variational upper bound for the ground-state energy, which converges to the exact value for $S\to 2^N$. Clearly, this limit is computationally impractical for system sizes $N\gg 10$. Yet, suitable criteria to select the subset of basis configurations lead to accurate approximations for feasible value of $S$.

In our framework, the selected configurations are sampled using autoregressive NNs, which can be trained to minimize $E$. 
%In this case, the energy to be minimized is the lowest eigenvalue of the subspace-Hamiltonian built on the configurations sampled by the ANN.
%Therefore, the neural network is trained to sample the bitstring configurations which provide the lowest eigenvalue. All the details about the loss functions and their gradients are in the Appendix~\ref{supp}. 
More precisely, as in Ref.~\cite{Robledo_Moreno_2025}, we define the loss function to be minimized during training as
\begin{equation}\label{loss}
    L = \sum_kP(S^{(k)})E^{(k)} \, ,
\end{equation}
where $S^{(k)}$ represents a batch of bitstrings and $E^{(k)}$ is the lowest eigenvalue of the subspace Hamiltonian built on the $k$-th batch of bitstrings. In our case, the probability $P(S^{(k)})$ of sampling $S^{(k)}$ is given by the autoregressive NN. The minimization is performed using stochastic gradient-based methods, and the derivation of the gradient of $L$ with respect to the weights of the NN is shown in the Appendix~\ref{app1}.
%Here, we just mention that, during training, the gradient of the loss function $L$ with respect to the generic trainable parameter $\omega$ of the neural network can be calculated as
%
%\begin{equation}
   % \frac{\partial L}{\partial\omega} = \frac{1}{K}\sum_{k=1}^K\biggl( \sum_{x^{(l)}\in S^{(k)}} \frac{\partial \log(P(x^{(l)}))}{\partial\omega} \biggr ) E^{(k)} \, ,
%\end{equation}
%
%where $K$ is the number of different batches $S^{(k)}$ of bitstrings calculated at each step, $P(x^{(l)})$ is the probability of a specific bitstring $x^{(l)}$ given by the ANN, and $E^{(k)}$ is the lowest eigenvalue of the subspace-Hamiltonian built using the bitstrings in $S^{(k)}$. 

The testbeds considered in this article are quantum Ising models described by the following Hamiltonian:
\begin{equation}\label{hamiltonian}
    H = -\sum_{\langle i,j \rangle}J_{ij} Z_iZ_{j} - h\sum_i X_i \, ,
\end{equation}
where $Z_i$ and $X_i$ are Pauli operators acting on spin $i$, $J_{ij}$ represents the interaction strength between the nearest-neighbor spins $i$ and $j$, and $h$ is the transverse field strength.
Specifically, we consider three variants of this model:
\begin{enumerate}
    \item 1D ferromagnetic transverse field Ising model (1D-TFIM), with $J_{ii+1}=1$ for $i=1,\dots,N$, and periodic boundary conditions, i.e., the spin $N+1$ is identified with the spin $1$;
    %\item 1D chain with random $J_{ij}\in[0,1]$, and periodic boundary conditions (model B);
    \item 2D ferromagnetic TFIM (2D-TFIM), with $J_{ij}=1$ for $i$ and $j$ nearest-neighbor spins on a square lattice, and open boundary conditions;
    \item 2D Edward-Anderson model (2D-EAM) on a square lattice, with $J_{ij}$ randomly sampled from a normal distribution $\mathcal{N}(0,1)$ with zero mean and unit variance, and periodic boundary conditions.
\end{enumerate}
It is worth pointing out that quantum Monte Carlo simulations of quantum Ising models are not affected by the negative sign problem. Thus, by adopting these models as testbeds for the SBD approaches, we have the opportunity to make comparisons against unbiased estimates of the ground-state energy, even beyond the 1D case where the Jordan-Wigner theory provides the exact solution. In particular, the 2D-EAM represents a challenging testbed due to the presence of disordered frustrated interactions.
In addition, tuning $h$ allows us to control the concentration of the ground state~\cite{yu2025quantumcentricalgorithmsamplebasedkrylov}, thus testing the SBD approaches in different regimes.
Indeed, for $h\to\infty$, the ground state of these models tends to an equally weighted superposition of the computational basis elements, i.e. $|++\dots+\rangle$ (with $|+\rangle=\frac{1}{\sqrt{2}}(|0\rangle+|1\rangle)$). 
%Consequently, in this regime, the ground state becomes localized in the 
%$|+\rangle,|-\rangle$ basis, and highly delocalized in the computational basis. 
%
In the opposite limit $h\to 0$, quantum fluctuations vanish, leading to a more concentrated ground-state wave function in the chosen computational basis. 
%
%Clearly, tuning $h$ allows us to control the concentration of the ground state~\cite{yu2025quantumcentricalgorithmsamplebasedkrylov}, thus testing the SBD approaches in different regimes.
It is worth emphasizing that generic SBD methods, including our SND method without adaptive basis changes, are expected to perform well only for relatively small $h$. Numerical results confirming this expectation are shown in the Appendix~\ref{app1}.

To extend the regime of applicability of SBD approaches, we introduce the AB-SND method. AB-SND improves the SND strategy by incorporating a basis transformation defined by a parameterized unitary operator $U(\boldsymbol{\theta})$, which maps the original computational basis to a rotated basis in which the ground state is more concentrated. As in SND, an autoregressive NN generates bitstring samples, but these are rotated by $U(\boldsymbol{\theta})$. The subspace Hamiltonian is then constructed from transformed matrix elements $\langle x^{(l)} | U(\boldsymbol{\theta})^\dag H U(\boldsymbol{\theta}) | x^{(m)} \rangle$, as shown in Fig.~\ref{fig1}. In most of our experiments, we use a combination of single-spin rotations $U(\boldsymbol{\theta})=\bigotimes_{i=1,\dots N}\,U_i(\theta_i)$, where
\begin{equation}\label{Ry}
    U_i(\theta_i)=\begin{bmatrix}
			\cos{\frac{\theta_i}{2}} & -\sin{\frac{\theta_i}{2}} \\
			\sin{\frac{\theta_i}{2}} & \cos{\frac{\theta_i}{2}} 
		\end{bmatrix}\, ,
\end{equation}
and each angle $\theta_i$ is an independent parameter for spin $i$.
Because the rotations act independently on each spin and the Hamiltonian is composed of local Pauli operators, we can efficiently compute the transformed Hamiltonian $U^{\dag}HU$ using classical hardware. In addition to single-spin rotations, we also implement non-overlapping two-spin rotations, which increase the expressive power of the basis transformation while remaining classically tractable.
The unitary operator is defined as a composition of $R_Y$, $R_{ZZ}$, and $R_X$ gates that act on each spin in the pair. The matrix representations of these gates are as follows:
\begin{equation}
\begin{aligned}
    R_Y(\alpha)&=\begin{bmatrix}
			\cos{\frac{\alpha}{2}} & -\sin{\frac{\alpha}{2}} \\
			\sin{\frac{\alpha}{2}} & \cos{\frac{\alpha}{2}} 
		\end{bmatrix}\, ,
    \\
    \\
    R_X(\beta) &= 
    \begin{bmatrix}
        \cos\frac{\beta}{2} & -i\sin\frac{\beta}{2} \\
        -i\sin\frac{\beta}{2} & \cos\frac{\beta}{2}
    \end{bmatrix},
    \\
    \\
    R_{ZZ}(\gamma) &= 
    \begin{bmatrix}
        e^{-i\frac{\gamma}{2}} & 0 & 0 & 0 \\
        0 & e^{i\frac{\gamma}{2}} & 0 & 0 \\
        0 & 0 & e^{i\frac{\gamma}{2}} & 0 \\
        0 & 0 & 0 & e^{-i\frac{\gamma}{2}}
    \end{bmatrix} \, .
\end{aligned}
\end{equation}
% 
%where $\boldsymbol{\theta}=\{\alpha,\beta,\gamma\}$ are the rotational parameters optimized as explained in Appendix~\ref{app3}. 
As mentioned above, the gates $R_{ZZ}(\gamma)$ act on non-overlapping spin pairs $\{0,1\}$, $\{2,3\}$, etc. 
%Moreover, rotations involving only few, non-overlapping spins can be implemented classically. Results for two-spins rotations are shown in the Appendix~\ref{supp}.
However, the AB-SND framework is compatible with more general, potentially strongly entangling basis transformations, which could be implemented on quantum hardware. The use of quantum circuits for evaluating subspace matrix elements is discussed in Sec.~\ref{qc}. For AB-SND, the optimization of the angle parameter $\boldsymbol{\theta}$ can be approached in different ways, as discussed in the Appendices~\ref{app2} and \ref{app5}.

\begin{figure*}
	\centering
	\includegraphics[width=\textwidth]{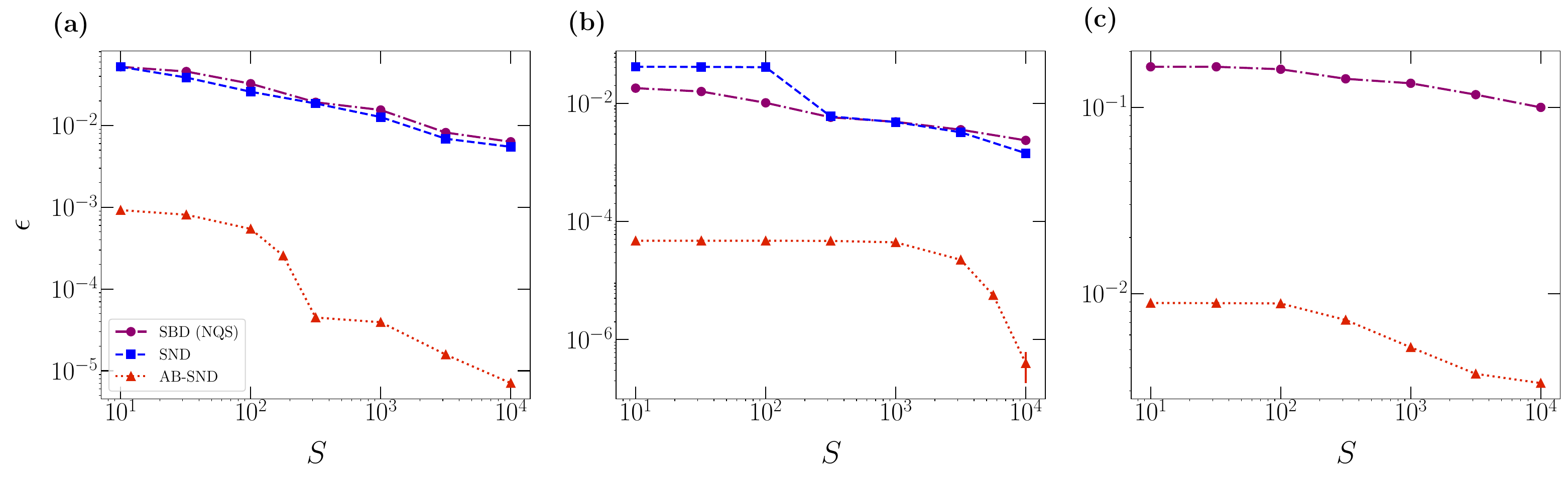}
	\caption{Relative error $\epsilon$ as a function of the number of unique configurations $S$ used to build the subspace Hamiltonian. Panels (a), (b), and (c) display the relative errors for the 1D-TFIM, the 2D-TFIM, and the 2D-EAM with $N=50$, $N=100$, and $N=64$ spins, respectively. The transverse field is $h=0.5$ for the 1D-TFIM and the 2D-TFIM, while it is $h=1$ for the 2D-EAM. In panel (b), we take into account the statistical uncertainty of the quantum Monte Carlo simulations used to determine the unbiased estimate of the ground-state energy.}
	\label{N50}
\end{figure*}

\section{Results}\label{results}
Hereafter, we analyze the performances of SND and AB-SND powered by local basis transformations on the three quantum spin models described in Sec.~\ref{methods}.
As a benchmark, we consider a more standard SBD approach in which the configurations are sampled from the (squared modulus) exact ground-state wave function or from a very accurate approximation obtained via a variational Monte Carlo simulation. It is worth emphasizing that this procedure assumes that an accurate representation of the ground state can be obtained through a complementary computational technique. This allows us to execute standard SBD under very favorable conditions, thus representing a stringent benchmark for novel SND approaches.
%Specifically, the exact ground state is obtained via exact diagonalization calculations for numbers of spins up to $N=25$. For larger system sizes, for which exact diagonalization is computationally impractical, 
We sample configurations from an NQS in the form of a restricted Boltzmann machine. The latter is optimized using the NetKet library~\cite{netket2:2019,netket3:2022}. 
Our numerical tests show that,  as a sampling engine for SBD approaches, the NQS ansatz performs essentially as well as the exact ground state, at least for the system sizes for which the latter can be computed. Thus, in the following, we mostly adopt NQS sampling, unless otherwise specified.
%

%
%Importantly, this is not a direct comparison with the VMC energy estimate, but rather we use the learned RBM state as a proxy for the ground state to sample configurations and construct a subspace-Hamiltonian via SBD. This sampling procedure is employed because we do not have access to the exact ground state distribution for large systems.

To quantify the performance of the various SBD approaches, we compute the relative error $\epsilon=|\frac{y-E_0}{E_0}|$, where $y$ represents the energy estimate from a given method, and $E_0$ is the exact ground-state energy. The latter is computed via the Jordan-Wigner transformations for the 1D model~\cite{PFEUTY197079,PhysRevB.53.8486}, while for the 2D models we employ continuous-time projection quantum Monte Carlo simulations~\cite{becca2017quantum, PhysRevE.110.065305}. The latter provide unbiased estimates, affected only by very small statistical uncertainties.

In Fig.~\ref{N50}, we analyze the relative errors in the three testbed models, considering relatively weak transverse fields $h$ for which the ground state is concentrated in the chosen computational basis.
As expected, all three SBD techniques perform well, showing a systematic accuracy improvement with the number of unique configurations $S$ used to build the subspace. However, the SND performance deteriorates for the 2D-EAM (not shown). This effect may be attributed to the rugged energy landscapes occurring in spin-glass phases~\cite{PhysRevResearch.2.023232}. 
In this testbed, the standard SBD method based on NQS-sampled configurations performs better, but still reaches errors as large as 10\% for computationally practical values of $S$. 
Notably, thanks to the additional variational flexibility introduced by the learnable basis transformation, the AB-SND method displays a systematic performance improvement with $S$ also in the 2D-EAM. In fact, it consistently outperforms the other SBD approaches we consider, in all three testbeds.
%We observe that for model C with $h=0.5$, we get better results than for the other 2 models. That's because, in that scenario, $h$ is far from the critical point. Similarly, if we employ AB-SND for model A with $h=0.25$, we get a relative error of $9\times10^{-7}$ with $S=10^4$ sampled spin configurations.
%SND or AB-SND may be an alternative to standard neural quantum states methods because they can potentially estimate the ground state energy using only approximate probabilities of the most important configurations. As a result, there's no need to learn the phases or signs of the basis elements, since these are recovered through diagonalization. \textcolor{blue}{Anche se SND senza cambiamento di base sembra non convergere in regimi non ferromagnetici, quindi per quella strategia non penso sia vero.} \textcolor{red}{For similar reasons, it is possible that quantum versions of SBD (SQD) can outperform noisy VQE implementations (quello che intendo è che la differenze tra SQD e VQE è simile alla differenza tra SND e NQS. In VQE e NQS si cerca di ottenere una forma variazionale dello stato, in SQD e SND invece hai bisogno semplicemente delle probabilità (neanche estremamente esatta in teoria) di ogni configurazione)} %Furthermore, SND results can be improved through a basis transformation, as demonstrated in AB-SND.

%Notably, the AB-SND technique is always outpeforming SND. 
%
\begin{figure*}
\centering
\includegraphics[width=\textwidth]{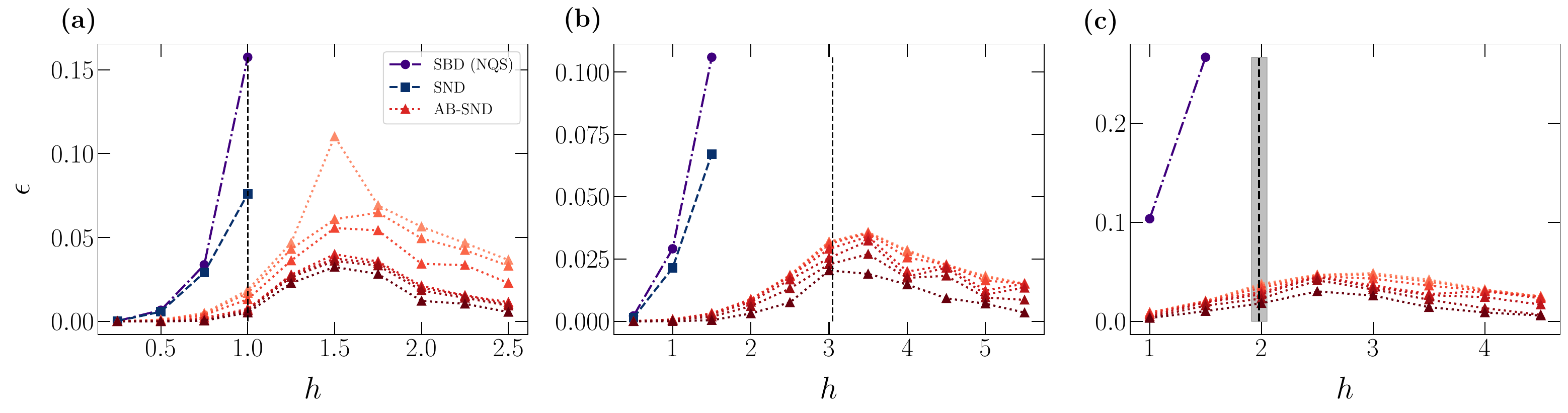}
\caption{Relative error $\epsilon$ as a function of the transverse field $h$, for different numbers of unique configurations $S$. For AB-SND, darker colors correspond to larger $S$, with $S\in\{10,10^{1.5},10^2,\dotsc,10^4\}$, while for SBD from the NQS and SND we consider only $S=10^4$.  Panel (a) represents the errors for the 1D-TFIM with $N=50$ spins, panel (b) corresponds to the 2D-TFIM with $N=100$ spins, and panel (c) to the 2D-EAM with $N=64$ spins. The vertical dashed lines represent the corresponding quantum critical points, namely, the ferromagnetic transitions in the 1D-TFIM and the 2D-TFIM, and the spin-glass transition in the 2D-EAM. For the latter, the uncertainty is represented by the gray bar~\cite{PhysRevE.110.065305}.}
\label{h}
\end{figure*}

In Fig.~\ref{h}, the relative energy error is plotted as a function of the transverse field $h$, for different numbers of unique configurations $S$. As expected, the performance of SND and standard NQS-based SBD methods rapidly deteriorates as $h$ increases, denoting the limited regime of applicability of these approaches. Instead, the AB-SND method, here implemented with single-spin rotations, is accurate also at significantly larger transverse fields. In fact, it reaches small relative errors also in the large $h$ limit.  This indicates that adaptive single-spin rotations enable a continuous interpolation between the original computational basis, in which the ground state is concentrated in the $h \to 0$ limit, and the basis built using eigenstates of the $X$ Pauli operator ${ |+ \rangle, | - \rangle }$, in which concentration occurs in the opposite limit. Sizable inaccuracies occur in the intermediate regime, approximately in the region around the quantum phase transitions occurring in the three testbed models, namely, the paramagnetic to ferromagnetic transition in the 1D-TFIM and the 2D-TFIM, and the quantum spin-glass transition in the 2D-EAM~\cite{PhysRevE.110.065305}.
In the Appendix~\ref{app4}, we provide numerical evidence that in the large $S$ limit, the peak of the energy error approaches the critical point $h_c = 1$ of the 1D-TFIM .

\begin{figure}
	\centering
	\includegraphics[width=\columnwidth]{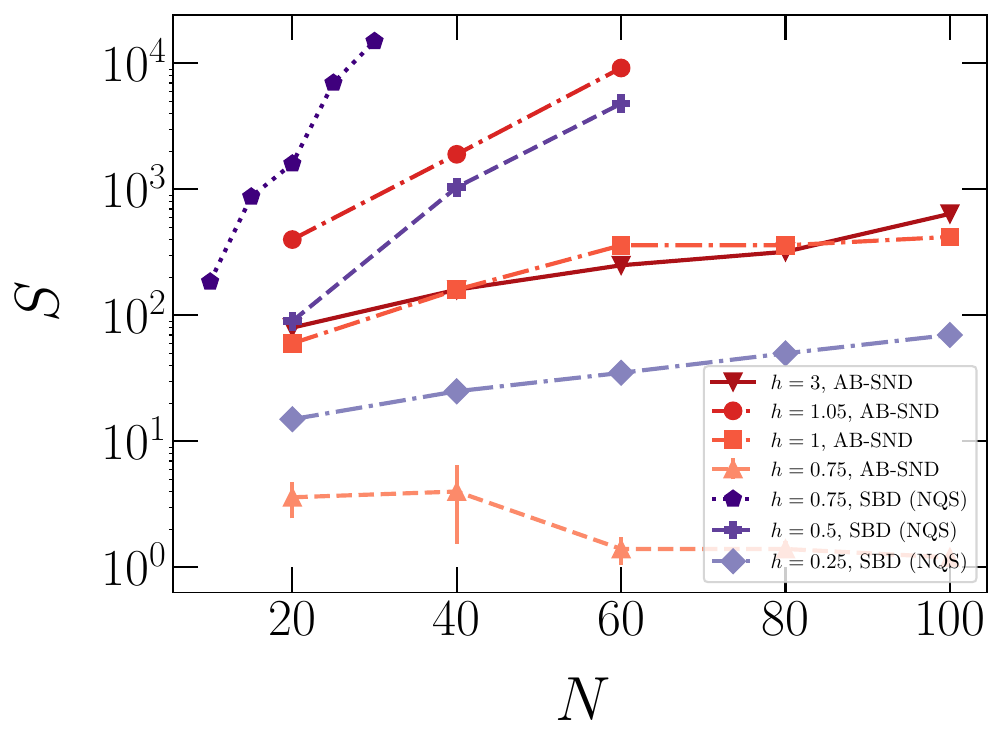}
	\caption{Number of unique configurations $S$ required to reach relative errors below $\epsilon=0.01$ as a function of the number of spins $N$ for the 1D-TFIM at different transverse fields $h$. We compare the performance of a standard SBD approach powered by NQS sampling (shades of blue) with the one of the AB-SND with single-spin rotations (shades of red). The error-bars for the AB-SND method for $h=0.5$ and $h=0.75$ represent the estimated standard deviation of the average over five training processes.} 
	\label{bdsnd_netket}
\end{figure}

In Fig.~\ref{bdsnd_netket}, we analyze how the computational cost scales with the system size $N$. Specifically, we determine the number of unique configurations $S$ required to reach a relative error of $1\%$, considering the 1D-TFIM. A standard SBD approach based on NQS sampling displays a problematic scaling already for $h \gtrsim 0.5$, making it impractical to reach system sizes $N \simeq 100 $ keeping the target accuracy. 
In this regime, the number of configurations required by the AB-SND approach powered by single-spin rotations is still essentially independent of $N$, denoting the important role of the basis change. However, the scaling approaches an exponential behavior slightly beyond the critical point $h_c=1$, while it improves again for transverse fields $h \gg 1$. 
Better accuracies in the critical regime $h \simeq 1$ can be obtained within the AB-SND approach with more general basis transformations, as discussed in the following.
%we show a comparison of the system size effect between AB-SND and standard SBD sampling from the NQS. For less challenging values of $h$ (as identified in Fig.\ref{h}), scaling to larger system sizes does not appear to significantly affect the performance. This is consistent with the observation that, when a suitable global rotation is applied, a single configuration can be effectively aligned with the ground state, resulting in near-perfect accuracy with  $S\simeq1$ across different system sizes. However, this effect is limited by the use of a single rotation, which becomes less effective slightly beyond the critical point, as introduced in Fig.\ref{h} and better displayed in Fig.~\ref{bdsnd_netket}. In these regimes, a negative impact of increasing system size on performance becomes evident. 
%

\begin{figure}
\centering
\includegraphics[width=\columnwidth]{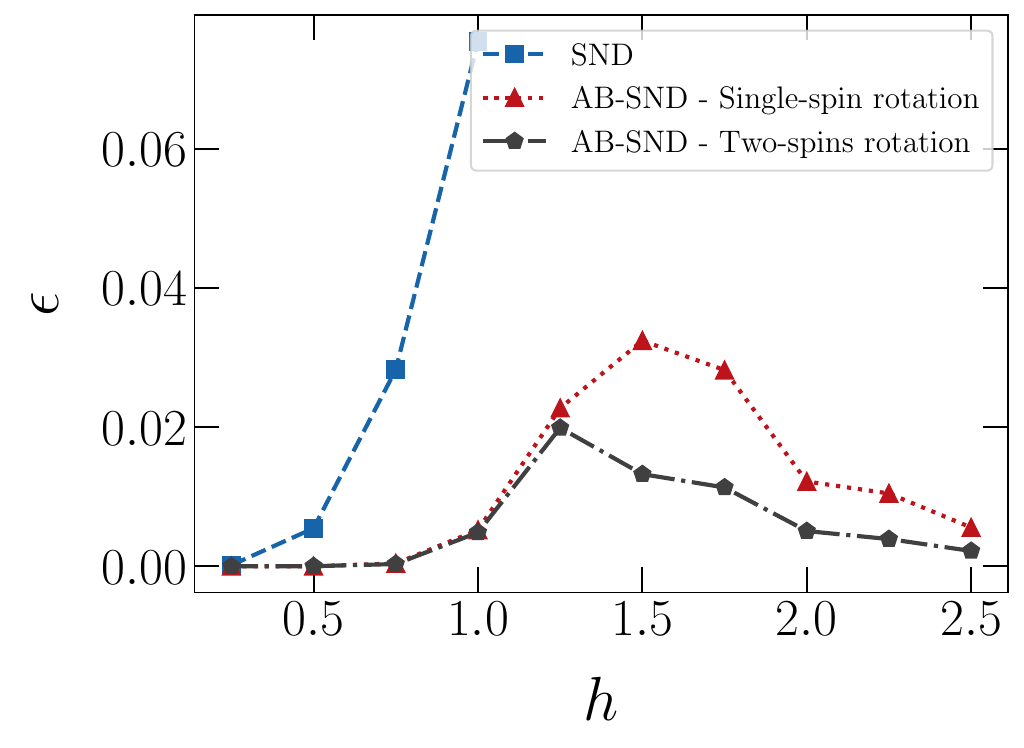}
\caption{Relative error $\epsilon$ as a function of transverse field $h$, for $S=10^4$ unique configurations in the truncated basis. The testbed is the 1D-TFIM with $N=50$ spins. We compare the performances of the SND approach, with the ones of AB-SND approaches featuring single-spin and non-overlapping two-spin rotations.}
\label{2q}
\end{figure}

In Fig.~\ref{2q}, we compare the accuracies of the AB-SND approaches powered by single-spin rotations and by non-overlapping two-spin unitary operators. 
The latter approach allows introducing some entanglement and provides a more expressive basis transformation, while remaining classically tractable.
In fact, we find that, at and slightly beyond the critical regime $h \gtrsim 1$, two-spin rotations lead to a sizable accuracy improvement compared to the single-spin case and, of course, compared to standard SBD based on the NQS-approximated ground-state sampling.
Even better accuracies can be obtained by implementing classically intractable basis transformations using quantum hardware, as discussed in Section~\ref{qc}.

An additional important challenge for all stochastic SBD methods is the decreasing efficiency of sampling unique configurations as the number of samples increases. This phenomenon, also noted in Ref.~\cite{reinholdt2025fundamentallimitationssamplebasedquantum}, can be addressed in our framework through effective temperature scaling during inference. This procedure is discussed in Appendix~\ref{app6}).

\section{Basis change using a quantum computer}\label{qc}
\begin{figure}
\centering
\includegraphics[width=\columnwidth]{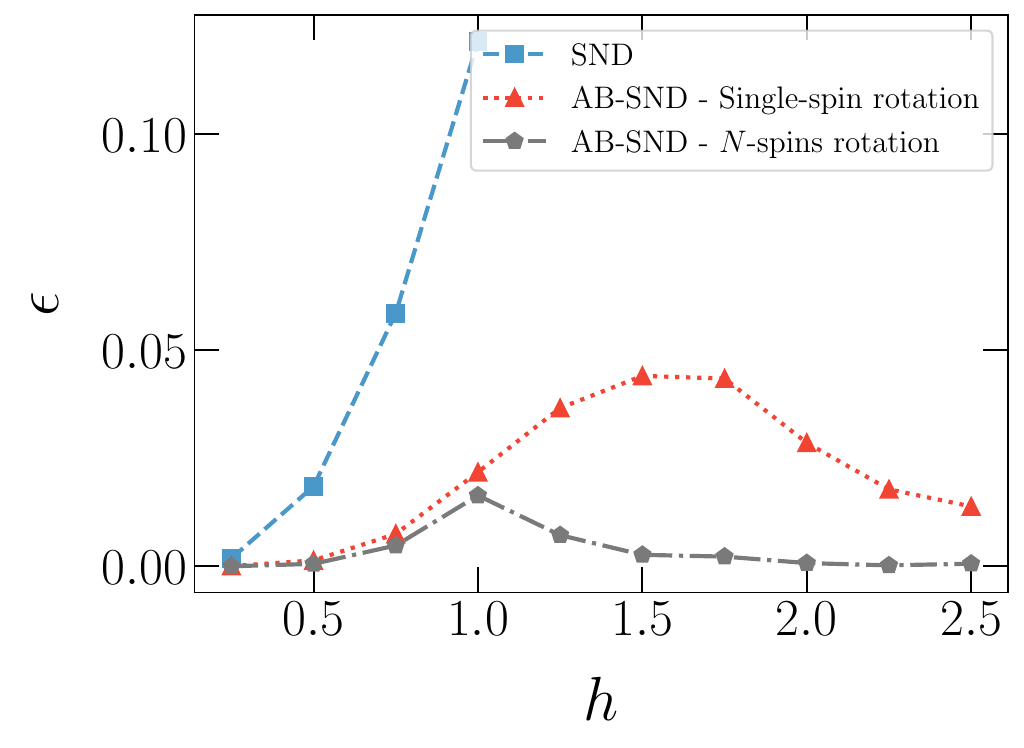}
\caption{Relative error $\epsilon$ as a function of the transverse field $h$, for $S=16$ unique configurations. The testbed model is the 1D-TFIM with $N=6$ spins. We compare the accuracies of the SND approach with the ones of the AB-SND methods with single-spin rotations and with a general basis change performed by the (simulated) quantum circuit described in the text.}
\label{Nq}
\end{figure}

For the quantum Ising models we consider, which feature up to two-spin couplings, basis changes based on combinations of single-spin and non-overlapping two-spin transformations can be efficiently performed on classical computers. 
More expressive basis changes could be efficiently performed using quantum circuits. Such transformations could further improve the performance of AB-SND approaches, especially near critical points. Indeed, a variational quantum eigensolver (VQE) algorithm has been shown to be able to accurately solve the 1D-TFIM also at criticality by optimizing a unitary transformation $U(\boldsymbol{\theta})$. This is achieved by minimizing the expectation value $\langle 0|U^\dag(\boldsymbol{\theta})HU(\boldsymbol{\theta})|0\rangle$~\cite{Ho_2019}, which is estimated using (typically large) shot numbers. This is  equivalent to a specific AB-SND strategy with only a single sampled configuration, namely, the state $|0\rangle=|00\dots0\rangle$. The general AB-SND  approach extends VQE by including the sampling of more basis elements, which leads to a finite matrix whose lowest eigenvalue is to be determined. 
Hereafter, we discuss how to implement a generic basis-change procedure using quantum computers. First, we set up a quantum circuit representing a parametrized basis-change unitary operator $U(\boldsymbol{\theta})$. Then, we can calculate the subspace-Hamiltonian elements $\langle x^{(l)}|U^\dag(\boldsymbol{\theta})HU(\boldsymbol{\theta})|x^{(m)}\rangle$ using the approach introduced in Ref.~\cite{D2SC05371C}. The diagonal terms $H_l =\langle x^{(l)}|U^\dag(\boldsymbol{\theta})HU(\boldsymbol{\theta})|x^{(l)}\rangle$ can be computed as standard expectation values. The off-diagonal terms can be calculated noticing that (using $i=\sqrt{-1}$)
\begin{equation}
    \Re\langle x^{(l)}|U^\dag(\boldsymbol{\theta})HU(\boldsymbol{\theta})|x^{(m)}\rangle = H_{l+m} - \frac{H_l}{2} - \frac{H_m}{2} \, ,
\end{equation}
and
\begin{equation}
    \Im\langle x^{(l)}|U^\dag(\boldsymbol{\theta})HU(\boldsymbol{\theta})|x^{(m)}\rangle = -H_{l+im} + \frac{H_l}{2} + \frac{H_m}{2} \, ,
\end{equation}
where 
\begin{equation}
    H_{l+m}= \frac{1}{\sqrt{2}}(\langle x^{(l)}|+\langle x^{(m)}|)U^\dag(\boldsymbol{\theta})HU(\boldsymbol{\theta})\frac{1}{\sqrt{2}}(|x^{(l)} \rangle+|x^{(m)}\rangle)
\end{equation}
and 
\begin{equation}
    H_{l+im}=\frac{1}{\sqrt{2}}(\langle x^{(l)}|-i\langle x^{(m)}|)U^\dag(\boldsymbol{\theta})HU(\boldsymbol{\theta})\frac{1}{\sqrt{2}}(|x^{(l)} \rangle+i|x^{(m)}\rangle) \, .
\end{equation}
Finally, in our framework, both the parameters $\boldsymbol{\theta}$ and the weights of the autoregressive NN from which the bitstrings $x^{(l)}$ are sampled can be optimized as explained for the AB-SND procedure with single-spin rotations. 
To test this approach, we implement a small numerical experiment using a classical simulation of quantum circuits with $N=6$ qubits for the 1D-TFIM. The circuit ansatz we choose is rather shallow. It features one layer of $R_Y$ gates acting on each qubit, and two blocks including $R_{ZZ}$ operators acting on all nearest-neighbor pairs and $R_X$ rotations acting on each qubit. 
As shown in Fig.~\ref{Nq}, the improvement of the AB-SND predictions based on the circuit-based basis change compared with the case of single-spin transformations is significant. With the more general transformation, appreciable inaccuracies occur only very close to the critical point $h_c=1$. We expect even better performances to be obtained by implementing basis transformations using deeper quantum circuits.

\section{Conclusions}\label{conclusion}
In this article, we introduced SND and its basis-adaptive extension, dubbed AB-SND. These are two NN-assisted sample-based diagonalization techniques for estimating ground-state energies of quantum many-body systems. Our tests show that AB-SND offers significant improvements over conventional SBD methods, particularly in regimes where the ground state is delocalized in the computational basis.

To extend the AB-SND approach beyond basis changes based on single-spin and two-spin operators, we explored the integration of AB-SND with quantum computing. As we explained, this allows the implementation of more general basis transformations. To test this procedure, we implemented a small-scale proof-of-concept experiment using classically simulated quantum circuits. In this setting, we used parametrized entangling gates to define a more expressive unitary $U(\boldsymbol{\theta})$, which led to improved performance compared to local spin rotations. Although these results are currently limited to small systems simulated classically, they suggest that AB-SND can be extended to hybrid quantum-classical workflows and could benefit from access to real quantum hardware in the future.
In fact, the AB-SND approach driven by quantum circuits represents an extension of the VQE algorithm beyond the case of a single initial state.

Further developments could include the use of more expressive basis-change circuits, improved optimization strategies in complex energy landscapes, and systematic explorations of the performance of AB-SND methods in larger or more strongly correlated systems. By combining neural sampling with learnable basis transformations, AB-SND provides a flexible and scalable framework for studying quantum many-body problems across a wide range of regimes.

The essential scripts used in this study are available on GitHub~\cite{Cantori_Adaptive-basis_sample-based_neural}.

\begin{acknowledgments}
%We are thankful to Emanuele Costa, Andrea Mari, Giuseppe Scriva, and David Vitali for fruitful discussions.
Support from the following sources is acknowledged: 
PNRR MUR project PE0000023-NQSTI; 
PRIN 2022 MUR project ``Hybrid algorithms for quantum simulators'' -- 2022H77XB7; 
PRIN-PNRR 2022 MUR project ``UEFA'' -- P2022NMBAJ; 
National Centre for HPC, Big Data and Quantum Computing (ICSC), CN00000013 Spoke 7 -- Materials \& Molecular Sciences, under project INNOVATOR; 
CINECA awards ISCRA  IsCc2\_REASON and INF25\_lincoln, for the availability of high-performance computing resources and support; 
EuroHPC Joint Undertaking for awarding access to the EuroHPC supercomputer LUMI, hosted by CSC (Finland), through EuroHPC Development and Regular Access calls. 
E.R. acknowledges support from the Catalonia Quantum Academy Student Research Fellowship 2024 for his research internship at the University of Barcelona.
\end{acknowledgments}

\bibliography{mybibliography}

\begin{thebibliography}{40}%
\makeatletter
\providecommand \@ifxundefined [1]{%
 \@ifx{#1\undefined}
}%
\providecommand \@ifnum [1]{%
 \ifnum #1\expandafter \@firstoftwo
 \else \expandafter \@secondoftwo
 \fi
}%
\providecommand \@ifx [1]{%
 \ifx #1\expandafter \@firstoftwo
 \else \expandafter \@secondoftwo
 \fi
}%
\providecommand \natexlab [1]{#1}%
\providecommand \enquote  [1]{``#1''}%
\providecommand \bibnamefont  [1]{#1}%
\providecommand \bibfnamefont [1]{#1}%
\providecommand \citenamefont [1]{#1}%
\providecommand \href@noop [0]{\@secondoftwo}%
\providecommand \href [0]{\begingroup \@sanitize@url \@href}%
\providecommand \@href[1]{\@@startlink{#1}\@@href}%
\providecommand \@@href[1]{\endgroup#1\@@endlink}%
\providecommand \@sanitize@url [0]{\catcode `\\12\catcode `\$12\catcode `\&12\catcode `\#12\catcode `\^12\catcode `\_12\catcode `\%12\relax}%
\providecommand \@@startlink[1]{}%
\providecommand \@@endlink[0]{}%
\providecommand \url  [0]{\begingroup\@sanitize@url \@url }%
\providecommand \@url [1]{\endgroup\@href {#1}{\urlprefix }}%
\providecommand \urlprefix  [0]{URL }%
\providecommand \Eprint [0]{\href }%
\providecommand \doibase [0]{https://doi.org/}%
\providecommand \selectlanguage [0]{\@gobble}%
\providecommand \bibinfo  [0]{\@secondoftwo}%
\providecommand \bibfield  [0]{\@secondoftwo}%
\providecommand \translation [1]{[#1]}%
\providecommand \BibitemOpen [0]{}%
\providecommand \bibitemStop [0]{}%
\providecommand \bibitemNoStop [0]{.\EOS\space}%
\providecommand \EOS [0]{\spacefactor3000\relax}%
\providecommand \BibitemShut  [1]{\csname bibitem#1\endcsname}%
\let\auto@bib@innerbib\@empty
%</preamble>
\bibitem [{\citenamefont {Carleo}\ \emph {et~al.}(2019{\natexlab{a}})\citenamefont {Carleo}, \citenamefont {Cirac}, \citenamefont {Cranmer}, \citenamefont {Daudet}, \citenamefont {Schuld}, \citenamefont {Tishby}, \citenamefont {Vogt-Maranto},\ and\ \citenamefont {Zdeborov\'a}}]{RevModPhys.91.045002}%
  \BibitemOpen
  \bibfield  {author} {\bibinfo {author} {\bibfnamefont {G.}~\bibnamefont {Carleo}}, \bibinfo {author} {\bibfnamefont {I.}~\bibnamefont {Cirac}}, \bibinfo {author} {\bibfnamefont {K.}~\bibnamefont {Cranmer}}, \bibinfo {author} {\bibfnamefont {L.}~\bibnamefont {Daudet}}, \bibinfo {author} {\bibfnamefont {M.}~\bibnamefont {Schuld}}, \bibinfo {author} {\bibfnamefont {N.}~\bibnamefont {Tishby}}, \bibinfo {author} {\bibfnamefont {L.}~\bibnamefont {Vogt-Maranto}},\ and\ \bibinfo {author} {\bibfnamefont {L.}~\bibnamefont {Zdeborov\'a}},\ }\bibfield  {title} {\bibinfo {title} {Machine learning and the physical sciences},\ }\href {https://doi.org/10.1103/RevModPhys.91.045002} {\bibfield  {journal} {\bibinfo  {journal} {Rev. Mod. Phys.}\ }\textbf {\bibinfo {volume} {91}},\ \bibinfo {pages} {045002} (\bibinfo {year} {2019}{\natexlab{a}})}\BibitemShut {NoStop}%
\bibitem [{\citenamefont {Carrasquilla}(2020)}]{Carrasquilla01012020}%
  \BibitemOpen
  \bibfield  {author} {\bibinfo {author} {\bibfnamefont {J.}~\bibnamefont {Carrasquilla}},\ }\bibfield  {title} {\bibinfo {title} {Machine learning for quantum matter},\ }\href {https://doi.org/10.1080/23746149.2020.1797528} {\bibfield  {journal} {\bibinfo  {journal} {Advances in Physics: X}\ }\textbf {\bibinfo {volume} {5}},\ \bibinfo {pages} {1797528} (\bibinfo {year} {2020})},\ \Eprint {https://arxiv.org/abs/https://doi.org/10.1080/23746149.2020.1797528} {https://doi.org/10.1080/23746149.2020.1797528} \BibitemShut {NoStop}%
\bibitem [{\citenamefont {Kulik}\ \emph {et~al.}(2022)\citenamefont {Kulik}, \citenamefont {Hammerschmidt}, \citenamefont {Schmidt}, \citenamefont {Botti}, \citenamefont {Marques}, \citenamefont {Boley}, \citenamefont {Scheffler}, \citenamefont {Todorović}, \citenamefont {Rinke}, \citenamefont {Oses}, \citenamefont {Smolyanyuk}, \citenamefont {Curtarolo}, \citenamefont {Tkatchenko}, \citenamefont {Bartók}, \citenamefont {Manzhos}, \citenamefont {Ihara}, \citenamefont {Carrington}, \citenamefont {Behler}, \citenamefont {Isayev}, \citenamefont {Veit}, \citenamefont {Grisafi}, \citenamefont {Nigam}, \citenamefont {Ceriotti}, \citenamefont {Schütt}, \citenamefont {Westermayr}, \citenamefont {Gastegger}, \citenamefont {Maurer}, \citenamefont {Kalita}, \citenamefont {Burke}, \citenamefont {Nagai}, \citenamefont {Akashi}, \citenamefont {Sugino}, \citenamefont {Hermann}, \citenamefont {Noé}, \citenamefont {Pilati}, \citenamefont {Draxl}, \citenamefont {Kuban}, \citenamefont {Rigamonti}, \citenamefont {Scheidgen},
  \citenamefont {Esters}, \citenamefont {Hicks}, \citenamefont {Toher}, \citenamefont {Balachandran}, \citenamefont {Tamblyn}, \citenamefont {Whitelam}, \citenamefont {Bellinger},\ and\ \citenamefont {Ghiringhelli}}]{Kulik_2022}%
  \BibitemOpen
  \bibfield  {author} {\bibinfo {author} {\bibfnamefont {H.~J.}\ \bibnamefont {Kulik}}, \bibinfo {author} {\bibfnamefont {T.}~\bibnamefont {Hammerschmidt}}, \bibinfo {author} {\bibfnamefont {J.}~\bibnamefont {Schmidt}}, \bibinfo {author} {\bibfnamefont {S.}~\bibnamefont {Botti}}, \bibinfo {author} {\bibfnamefont {M.~A.~L.}\ \bibnamefont {Marques}}, \bibinfo {author} {\bibfnamefont {M.}~\bibnamefont {Boley}}, \bibinfo {author} {\bibfnamefont {M.}~\bibnamefont {Scheffler}}, \bibinfo {author} {\bibfnamefont {M.}~\bibnamefont {Todorović}}, \bibinfo {author} {\bibfnamefont {P.}~\bibnamefont {Rinke}}, \bibinfo {author} {\bibfnamefont {C.}~\bibnamefont {Oses}}, \bibinfo {author} {\bibfnamefont {A.}~\bibnamefont {Smolyanyuk}}, \bibinfo {author} {\bibfnamefont {S.}~\bibnamefont {Curtarolo}}, \bibinfo {author} {\bibfnamefont {A.}~\bibnamefont {Tkatchenko}}, \bibinfo {author} {\bibfnamefont {A.~P.}\ \bibnamefont {Bartók}}, \bibinfo {author} {\bibfnamefont {S.}~\bibnamefont {Manzhos}}, \bibinfo {author} {\bibfnamefont
  {M.}~\bibnamefont {Ihara}}, \bibinfo {author} {\bibfnamefont {T.}~\bibnamefont {Carrington}}, \bibinfo {author} {\bibfnamefont {J.}~\bibnamefont {Behler}}, \bibinfo {author} {\bibfnamefont {O.}~\bibnamefont {Isayev}}, \bibinfo {author} {\bibfnamefont {M.}~\bibnamefont {Veit}}, \bibinfo {author} {\bibfnamefont {A.}~\bibnamefont {Grisafi}}, \bibinfo {author} {\bibfnamefont {J.}~\bibnamefont {Nigam}}, \bibinfo {author} {\bibfnamefont {M.}~\bibnamefont {Ceriotti}}, \bibinfo {author} {\bibfnamefont {K.~T.}\ \bibnamefont {Schütt}}, \bibinfo {author} {\bibfnamefont {J.}~\bibnamefont {Westermayr}}, \bibinfo {author} {\bibfnamefont {M.}~\bibnamefont {Gastegger}}, \bibinfo {author} {\bibfnamefont {R.~J.}\ \bibnamefont {Maurer}}, \bibinfo {author} {\bibfnamefont {B.}~\bibnamefont {Kalita}}, \bibinfo {author} {\bibfnamefont {K.}~\bibnamefont {Burke}}, \bibinfo {author} {\bibfnamefont {R.}~\bibnamefont {Nagai}}, \bibinfo {author} {\bibfnamefont {R.}~\bibnamefont {Akashi}}, \bibinfo {author} {\bibfnamefont
  {O.}~\bibnamefont {Sugino}}, \bibinfo {author} {\bibfnamefont {J.}~\bibnamefont {Hermann}}, \bibinfo {author} {\bibfnamefont {F.}~\bibnamefont {Noé}}, \bibinfo {author} {\bibfnamefont {S.}~\bibnamefont {Pilati}}, \bibinfo {author} {\bibfnamefont {C.}~\bibnamefont {Draxl}}, \bibinfo {author} {\bibfnamefont {M.}~\bibnamefont {Kuban}}, \bibinfo {author} {\bibfnamefont {S.}~\bibnamefont {Rigamonti}}, \bibinfo {author} {\bibfnamefont {M.}~\bibnamefont {Scheidgen}}, \bibinfo {author} {\bibfnamefont {M.}~\bibnamefont {Esters}}, \bibinfo {author} {\bibfnamefont {D.}~\bibnamefont {Hicks}}, \bibinfo {author} {\bibfnamefont {C.}~\bibnamefont {Toher}}, \bibinfo {author} {\bibfnamefont {P.~V.}\ \bibnamefont {Balachandran}}, \bibinfo {author} {\bibfnamefont {I.}~\bibnamefont {Tamblyn}}, \bibinfo {author} {\bibfnamefont {S.}~\bibnamefont {Whitelam}}, \bibinfo {author} {\bibfnamefont {C.}~\bibnamefont {Bellinger}},\ and\ \bibinfo {author} {\bibfnamefont {L.~M.}\ \bibnamefont {Ghiringhelli}},\ }\bibfield  {title} {\bibinfo
  {title} {Roadmap on machine learning in electronic structure},\ }\href {https://doi.org/10.1088/2516-1075/ac572f} {\bibfield  {journal} {\bibinfo  {journal} {Electronic Structure}\ }\textbf {\bibinfo {volume} {4}},\ \bibinfo {pages} {023004} (\bibinfo {year} {2022})}\BibitemShut {NoStop}%
\bibitem [{\citenamefont {Dunjko}\ and\ \citenamefont {Briegel}(2018)}]{Dunjko_2018}%
  \BibitemOpen
  \bibfield  {author} {\bibinfo {author} {\bibfnamefont {V.}~\bibnamefont {Dunjko}}\ and\ \bibinfo {author} {\bibfnamefont {H.~J.}\ \bibnamefont {Briegel}},\ }\bibfield  {title} {\bibinfo {title} {Machine learning \& artificial intelligence in the quantum domain: a review of recent progress},\ }\href {https://doi.org/10.1088/1361-6633/aab406} {\bibfield  {journal} {\bibinfo  {journal} {Reports on Progress in Physics}\ }\textbf {\bibinfo {volume} {81}},\ \bibinfo {pages} {074001} (\bibinfo {year} {2018})}\BibitemShut {NoStop}%
\bibitem [{\citenamefont {Saraceni}\ \emph {et~al.}(2020)\citenamefont {Saraceni}, \citenamefont {Cantori},\ and\ \citenamefont {Pilati}}]{PhysRevE.102.033301}%
  \BibitemOpen
  \bibfield  {author} {\bibinfo {author} {\bibfnamefont {N.}~\bibnamefont {Saraceni}}, \bibinfo {author} {\bibfnamefont {S.}~\bibnamefont {Cantori}},\ and\ \bibinfo {author} {\bibfnamefont {S.}~\bibnamefont {Pilati}},\ }\bibfield  {title} {\bibinfo {title} {Scalable neural networks for the efficient learning of disordered quantum systems},\ }\href {https://doi.org/10.1103/PhysRevE.102.033301} {\bibfield  {journal} {\bibinfo  {journal} {Phys. Rev. E}\ }\textbf {\bibinfo {volume} {102}},\ \bibinfo {pages} {033301} (\bibinfo {year} {2020})}\BibitemShut {NoStop}%
\bibitem [{\citenamefont {Mills}\ \emph {et~al.}(2019)\citenamefont {Mills}, \citenamefont {Ryczko}, \citenamefont {Luchak}, \citenamefont {Domurad}, \citenamefont {Beeler},\ and\ \citenamefont {Tamblyn}}]{C8SC04578J}%
  \BibitemOpen
  \bibfield  {author} {\bibinfo {author} {\bibfnamefont {K.}~\bibnamefont {Mills}}, \bibinfo {author} {\bibfnamefont {K.}~\bibnamefont {Ryczko}}, \bibinfo {author} {\bibfnamefont {I.}~\bibnamefont {Luchak}}, \bibinfo {author} {\bibfnamefont {A.}~\bibnamefont {Domurad}}, \bibinfo {author} {\bibfnamefont {C.}~\bibnamefont {Beeler}},\ and\ \bibinfo {author} {\bibfnamefont {I.}~\bibnamefont {Tamblyn}},\ }\bibfield  {title} {\bibinfo {title} {Extensive deep neural networks for transferring small scale learning to large scale systems},\ }\href {https://doi.org/10.1039/C8SC04578J} {\bibfield  {journal} {\bibinfo  {journal} {Chem. Sci.}\ }\textbf {\bibinfo {volume} {10}},\ \bibinfo {pages} {4129} (\bibinfo {year} {2019})}\BibitemShut {NoStop}%
\bibitem [{\citenamefont {Faber}\ \emph {et~al.}(2017)\citenamefont {Faber}, \citenamefont {Hutchison}, \citenamefont {Huang}, \citenamefont {Gilmer}, \citenamefont {Schoenholz}, \citenamefont {Dahl}, \citenamefont {Vinyals}, \citenamefont {Kearnes}, \citenamefont {Riley},\ and\ \citenamefont {von Lilienfeld}}]{doi:10.1021/acs.jctc.7b00577}%
  \BibitemOpen
  \bibfield  {author} {\bibinfo {author} {\bibfnamefont {F.~A.}\ \bibnamefont {Faber}}, \bibinfo {author} {\bibfnamefont {L.}~\bibnamefont {Hutchison}}, \bibinfo {author} {\bibfnamefont {B.}~\bibnamefont {Huang}}, \bibinfo {author} {\bibfnamefont {J.}~\bibnamefont {Gilmer}}, \bibinfo {author} {\bibfnamefont {S.~S.}\ \bibnamefont {Schoenholz}}, \bibinfo {author} {\bibfnamefont {G.~E.}\ \bibnamefont {Dahl}}, \bibinfo {author} {\bibfnamefont {O.}~\bibnamefont {Vinyals}}, \bibinfo {author} {\bibfnamefont {S.}~\bibnamefont {Kearnes}}, \bibinfo {author} {\bibfnamefont {P.~F.}\ \bibnamefont {Riley}},\ and\ \bibinfo {author} {\bibfnamefont {O.~A.}\ \bibnamefont {von Lilienfeld}},\ }\bibfield  {title} {\bibinfo {title} {Prediction errors of molecular machine learning models lower than hybrid {DFT} error},\ }\href {https://doi.org/10.1021/acs.jctc.7b00577} {\bibfield  {journal} {\bibinfo  {journal} {Journal of Chemical Theory and Computation}\ }\textbf {\bibinfo {volume} {13}},\ \bibinfo {pages} {5255} (\bibinfo {year}
  {2017})},\ \bibinfo {note} {pMID: 28926232},\ \Eprint {https://arxiv.org/abs/https://doi.org/10.1021/acs.jctc.7b00577} {https://doi.org/10.1021/acs.jctc.7b00577} \BibitemShut {NoStop}%
\bibitem [{\citenamefont {Carleo}\ and\ \citenamefont {Troyer}(2017)}]{Carleo_2017}%
  \BibitemOpen
  \bibfield  {author} {\bibinfo {author} {\bibfnamefont {G.}~\bibnamefont {Carleo}}\ and\ \bibinfo {author} {\bibfnamefont {M.}~\bibnamefont {Troyer}},\ }\bibfield  {title} {\bibinfo {title} {Solving the quantum many-body problem with artificial neural networks},\ }\href {https://doi.org/10.1126/science.aag2302} {\bibfield  {journal} {\bibinfo  {journal} {Science}\ }\textbf {\bibinfo {volume} {355}},\ \bibinfo {pages} {602–606} (\bibinfo {year} {2017})}\BibitemShut {NoStop}%
\bibitem [{\citenamefont {Sharir}\ \emph {et~al.}(2020)\citenamefont {Sharir}, \citenamefont {Levine}, \citenamefont {Wies}, \citenamefont {Carleo},\ and\ \citenamefont {Shashua}}]{PhysRevLett.124.020503}%
  \BibitemOpen
  \bibfield  {author} {\bibinfo {author} {\bibfnamefont {O.}~\bibnamefont {Sharir}}, \bibinfo {author} {\bibfnamefont {Y.}~\bibnamefont {Levine}}, \bibinfo {author} {\bibfnamefont {N.}~\bibnamefont {Wies}}, \bibinfo {author} {\bibfnamefont {G.}~\bibnamefont {Carleo}},\ and\ \bibinfo {author} {\bibfnamefont {A.}~\bibnamefont {Shashua}},\ }\bibfield  {title} {\bibinfo {title} {Deep autoregressive models for the efficient variational simulation of many-body quantum systems},\ }\href {https://doi.org/10.1103/PhysRevLett.124.020503} {\bibfield  {journal} {\bibinfo  {journal} {Phys. Rev. Lett.}\ }\textbf {\bibinfo {volume} {124}},\ \bibinfo {pages} {020503} (\bibinfo {year} {2020})}\BibitemShut {NoStop}%
\bibitem [{\citenamefont {Barrett}\ \emph {et~al.}(2022)\citenamefont {Barrett}, \citenamefont {Malyshev},\ and\ \citenamefont {Lvovsky}}]{Barrett2022}%
  \BibitemOpen
  \bibfield  {author} {\bibinfo {author} {\bibfnamefont {T.~D.}\ \bibnamefont {Barrett}}, \bibinfo {author} {\bibfnamefont {A.}~\bibnamefont {Malyshev}},\ and\ \bibinfo {author} {\bibfnamefont {A.~I.}\ \bibnamefont {Lvovsky}},\ }\bibfield  {title} {\bibinfo {title} {Autoregressive neural-network wavefunctions for ab initio quantum chemistry},\ }\href {https://doi.org/10.1038/s42256-022-00461-z} {\bibfield  {journal} {\bibinfo  {journal} {Nature Machine Intelligence}\ }\textbf {\bibinfo {volume} {4}},\ \bibinfo {pages} {351} (\bibinfo {year} {2022})}\BibitemShut {NoStop}%
\bibitem [{\citenamefont {Abraham}\ and\ \citenamefont {Mayhall}(2020)}]{Abraham_2020}%
  \BibitemOpen
  \bibfield  {author} {\bibinfo {author} {\bibfnamefont {V.}~\bibnamefont {Abraham}}\ and\ \bibinfo {author} {\bibfnamefont {N.~J.}\ \bibnamefont {Mayhall}},\ }\bibfield  {title} {\bibinfo {title} {Selected configuration interaction in a basis of cluster state tensor products},\ }\href {https://doi.org/10.1021/acs.jctc.0c00141} {\bibfield  {journal} {\bibinfo  {journal} {Journal of Chemical Theory and Computation}\ }\textbf {\bibinfo {volume} {16}},\ \bibinfo {pages} {6098–6113} (\bibinfo {year} {2020})}\BibitemShut {NoStop}%
\bibitem [{\citenamefont {Craciunescu}\ \emph {et~al.}(2025)\citenamefont {Craciunescu}, \citenamefont {Prentice},\ and\ \citenamefont {Paterson}}]{10.1063/5.0233542}%
  \BibitemOpen
  \bibfield  {author} {\bibinfo {author} {\bibfnamefont {L.}~\bibnamefont {Craciunescu}}, \bibinfo {author} {\bibfnamefont {A.~W.}\ \bibnamefont {Prentice}},\ and\ \bibinfo {author} {\bibfnamefont {M.~J.}\ \bibnamefont {Paterson}},\ }\bibfield  {title} {\bibinfo {title} {Selected configuration interaction for high accuracy and compact wave functions: {Propane} as a case study},\ }\href {https://doi.org/10.1063/5.0233542} {\bibfield  {journal} {\bibinfo  {journal} {The Journal of Chemical Physics}\ }\textbf {\bibinfo {volume} {162}},\ \bibinfo {pages} {034102} (\bibinfo {year} {2025})},\ \Eprint {https://arxiv.org/abs/https://pubs.aip.org/aip/jcp/article-pdf/doi/10.1063/5.0233542/20349175/034102\_1\_5.0233542.pdf} {https://pubs.aip.org/aip/jcp/article-pdf/doi/10.1063/5.0233542/20349175/034102\_1\_5.0233542.pdf} \BibitemShut {NoStop}%
\bibitem [{\citenamefont {Giner}\ \emph {et~al.}(2013)\citenamefont {Giner}, \citenamefont {Scemama},\ and\ \citenamefont {Caffarel}}]{doi:10.1139/cjc-2013-0017}%
  \BibitemOpen
  \bibfield  {author} {\bibinfo {author} {\bibfnamefont {E.}~\bibnamefont {Giner}}, \bibinfo {author} {\bibfnamefont {A.}~\bibnamefont {Scemama}},\ and\ \bibinfo {author} {\bibfnamefont {M.}~\bibnamefont {Caffarel}},\ }\bibfield  {title} {\bibinfo {title} {Using perturbatively selected configuration interaction in quantum {Monte Carlo} calculations},\ }\href {https://doi.org/10.1139/cjc-2013-0017} {\bibfield  {journal} {\bibinfo  {journal} {Canadian Journal of Chemistry}\ }\textbf {\bibinfo {volume} {91}},\ \bibinfo {pages} {879} (\bibinfo {year} {2013})},\ \Eprint {https://arxiv.org/abs/https://doi.org/10.1139/cjc-2013-0017} {https://doi.org/10.1139/cjc-2013-0017} \BibitemShut {NoStop}%
\bibitem [{\citenamefont {Yanagisawa}(2007)}]{PhysRevB.75.224503}%
  \BibitemOpen
  \bibfield  {author} {\bibinfo {author} {\bibfnamefont {T.}~\bibnamefont {Yanagisawa}},\ }\bibfield  {title} {\bibinfo {title} {Quantum {Monte Carlo} diagonalization for many-fermion systems},\ }\href {https://doi.org/10.1103/PhysRevB.75.224503} {\bibfield  {journal} {\bibinfo  {journal} {Phys. Rev. B}\ }\textbf {\bibinfo {volume} {75}},\ \bibinfo {pages} {224503} (\bibinfo {year} {2007})}\BibitemShut {NoStop}%
\bibitem [{\citenamefont {Robledo-Moreno}\ \emph {et~al.}(2025)\citenamefont {Robledo-Moreno}, \citenamefont {Motta}, \citenamefont {Haas}, \citenamefont {Javadi-Abhari}, \citenamefont {Jurcevic}, \citenamefont {Kirby}, \citenamefont {Martiel}, \citenamefont {Sharma}, \citenamefont {Sharma}, \citenamefont {Shirakawa}, \citenamefont {Sitdikov}, \citenamefont {Sun}, \citenamefont {Sung}, \citenamefont {Takita}, \citenamefont {Tran}, \citenamefont {Yunoki},\ and\ \citenamefont {Mezzacapo}}]{Robledo_Moreno_2025}%
  \BibitemOpen
  \bibfield  {author} {\bibinfo {author} {\bibfnamefont {J.}~\bibnamefont {Robledo-Moreno}}, \bibinfo {author} {\bibfnamefont {M.}~\bibnamefont {Motta}}, \bibinfo {author} {\bibfnamefont {H.}~\bibnamefont {Haas}}, \bibinfo {author} {\bibfnamefont {A.}~\bibnamefont {Javadi-Abhari}}, \bibinfo {author} {\bibfnamefont {P.}~\bibnamefont {Jurcevic}}, \bibinfo {author} {\bibfnamefont {W.}~\bibnamefont {Kirby}}, \bibinfo {author} {\bibfnamefont {S.}~\bibnamefont {Martiel}}, \bibinfo {author} {\bibfnamefont {K.}~\bibnamefont {Sharma}}, \bibinfo {author} {\bibfnamefont {S.}~\bibnamefont {Sharma}}, \bibinfo {author} {\bibfnamefont {T.}~\bibnamefont {Shirakawa}}, \bibinfo {author} {\bibfnamefont {I.}~\bibnamefont {Sitdikov}}, \bibinfo {author} {\bibfnamefont {R.-Y.}\ \bibnamefont {Sun}}, \bibinfo {author} {\bibfnamefont {K.~J.}\ \bibnamefont {Sung}}, \bibinfo {author} {\bibfnamefont {M.}~\bibnamefont {Takita}}, \bibinfo {author} {\bibfnamefont {M.~C.}\ \bibnamefont {Tran}}, \bibinfo {author} {\bibfnamefont
  {S.}~\bibnamefont {Yunoki}},\ and\ \bibinfo {author} {\bibfnamefont {A.}~\bibnamefont {Mezzacapo}},\ }\bibfield  {title} {\bibinfo {title} {Chemistry beyond the scale of exact diagonalization on a quantum-centric supercomputer},\ }\bibfield  {journal} {\bibinfo  {journal} {Science Advances}\ }\textbf {\bibinfo {volume} {11}},\ \href {https://doi.org/10.1126/sciadv.adu9991} {10.1126/sciadv.adu9991} (\bibinfo {year} {2025})\BibitemShut {NoStop}%
\bibitem [{\citenamefont {Yoshioka}\ \emph {et~al.}(2025)\citenamefont {Yoshioka}, \citenamefont {Amico}, \citenamefont {Kirby}, \citenamefont {Jurcevic}, \citenamefont {Dutt}, \citenamefont {Fuller}, \citenamefont {Garion}, \citenamefont {Haas}, \citenamefont {Hamamura}, \citenamefont {Ivrii}, \citenamefont {Majumdar}, \citenamefont {Minev}, \citenamefont {Motta}, \citenamefont {Pokharel}, \citenamefont {Rivero}, \citenamefont {Sharma}, \citenamefont {Wood}, \citenamefont {Javadi-Abhari},\ and\ \citenamefont {Mezzacapo}}]{Yoshioka_2025}%
  \BibitemOpen
  \bibfield  {author} {\bibinfo {author} {\bibfnamefont {N.}~\bibnamefont {Yoshioka}}, \bibinfo {author} {\bibfnamefont {M.}~\bibnamefont {Amico}}, \bibinfo {author} {\bibfnamefont {W.}~\bibnamefont {Kirby}}, \bibinfo {author} {\bibfnamefont {P.}~\bibnamefont {Jurcevic}}, \bibinfo {author} {\bibfnamefont {A.}~\bibnamefont {Dutt}}, \bibinfo {author} {\bibfnamefont {B.}~\bibnamefont {Fuller}}, \bibinfo {author} {\bibfnamefont {S.}~\bibnamefont {Garion}}, \bibinfo {author} {\bibfnamefont {H.}~\bibnamefont {Haas}}, \bibinfo {author} {\bibfnamefont {I.}~\bibnamefont {Hamamura}}, \bibinfo {author} {\bibfnamefont {A.}~\bibnamefont {Ivrii}}, \bibinfo {author} {\bibfnamefont {R.}~\bibnamefont {Majumdar}}, \bibinfo {author} {\bibfnamefont {Z.}~\bibnamefont {Minev}}, \bibinfo {author} {\bibfnamefont {M.}~\bibnamefont {Motta}}, \bibinfo {author} {\bibfnamefont {B.}~\bibnamefont {Pokharel}}, \bibinfo {author} {\bibfnamefont {P.}~\bibnamefont {Rivero}}, \bibinfo {author} {\bibfnamefont {K.}~\bibnamefont {Sharma}}, \bibinfo
  {author} {\bibfnamefont {C.~J.}\ \bibnamefont {Wood}}, \bibinfo {author} {\bibfnamefont {A.}~\bibnamefont {Javadi-Abhari}},\ and\ \bibinfo {author} {\bibfnamefont {A.}~\bibnamefont {Mezzacapo}},\ }\bibfield  {title} {\bibinfo {title} {Krylov diagonalization of large many-body hamiltonians on a quantum processor},\ }\bibfield  {journal} {\bibinfo  {journal} {Nature Communications}\ }\textbf {\bibinfo {volume} {16}},\ \href {https://doi.org/10.1038/s41467-025-59716-z} {10.1038/s41467-025-59716-z} (\bibinfo {year} {2025})\BibitemShut {NoStop}%
\bibitem [{\citenamefont {Kanno}\ \emph {et~al.}(2023)\citenamefont {Kanno}, \citenamefont {Kohda}, \citenamefont {Imai}, \citenamefont {Koh}, \citenamefont {Mitarai}, \citenamefont {Mizukami},\ and\ \citenamefont {Nakagawa}}]{kanno2023quantumselectedconfigurationinteractionclassical}%
  \BibitemOpen
  \bibfield  {author} {\bibinfo {author} {\bibfnamefont {K.}~\bibnamefont {Kanno}}, \bibinfo {author} {\bibfnamefont {M.}~\bibnamefont {Kohda}}, \bibinfo {author} {\bibfnamefont {R.}~\bibnamefont {Imai}}, \bibinfo {author} {\bibfnamefont {S.}~\bibnamefont {Koh}}, \bibinfo {author} {\bibfnamefont {K.}~\bibnamefont {Mitarai}}, \bibinfo {author} {\bibfnamefont {W.}~\bibnamefont {Mizukami}},\ and\ \bibinfo {author} {\bibfnamefont {Y.~O.}\ \bibnamefont {Nakagawa}},\ }\href {https://arxiv.org/abs/2302.11320} {\bibinfo {title} {Quantum-selected configuration interaction: {Classical} diagonalization of {Hamiltonians} in subspaces selected by quantum computers}} (\bibinfo {year} {2023}),\ \Eprint {https://arxiv.org/abs/2302.11320} {arXiv:2302.11320 [quant-ph]} \BibitemShut {NoStop}%
\bibitem [{\citenamefont {Arute}\ \emph {et~al.}(2019)\citenamefont {Arute}, \citenamefont {Arya}, \citenamefont {Babbush}, \citenamefont {Bacon}, \citenamefont {Bardin}, \citenamefont {Barends}, \citenamefont {Biswas}, \citenamefont {Boixo}, \citenamefont {Brandao}, \citenamefont {Buell} \emph {et~al.}}]{arute2019quantum}%
  \BibitemOpen
  \bibfield  {author} {\bibinfo {author} {\bibfnamefont {F.}~\bibnamefont {Arute}}, \bibinfo {author} {\bibfnamefont {K.}~\bibnamefont {Arya}}, \bibinfo {author} {\bibfnamefont {R.}~\bibnamefont {Babbush}}, \bibinfo {author} {\bibfnamefont {D.}~\bibnamefont {Bacon}}, \bibinfo {author} {\bibfnamefont {J.~C.}\ \bibnamefont {Bardin}}, \bibinfo {author} {\bibfnamefont {R.}~\bibnamefont {Barends}}, \bibinfo {author} {\bibfnamefont {R.}~\bibnamefont {Biswas}}, \bibinfo {author} {\bibfnamefont {S.}~\bibnamefont {Boixo}}, \bibinfo {author} {\bibfnamefont {F.~G.}\ \bibnamefont {Brandao}}, \bibinfo {author} {\bibfnamefont {D.~A.}\ \bibnamefont {Buell}}, \emph {et~al.},\ }\bibfield  {title} {\bibinfo {title} {Quantum supremacy using a programmable superconducting processor},\ }\href {https://doi.org/https://doi.org/10.1126/sciadv.adu9991} {\bibfield  {journal} {\bibinfo  {journal} {Nature}\ }\textbf {\bibinfo {volume} {574}},\ \bibinfo {pages} {505} (\bibinfo {year} {2019})}\BibitemShut {NoStop}%
\bibitem [{\citenamefont {Coe}(2018)}]{Coe_2018}%
  \BibitemOpen
  \bibfield  {author} {\bibinfo {author} {\bibfnamefont {J.~P.}\ \bibnamefont {Coe}},\ }\bibfield  {title} {\bibinfo {title} {Machine learning configuration interaction},\ }\href {https://doi.org/10.1021/acs.jctc.8b00849} {\bibfield  {journal} {\bibinfo  {journal} {Journal of Chemical Theory and Computation}\ }\textbf {\bibinfo {volume} {14}},\ \bibinfo {pages} {5739–5749} (\bibinfo {year} {2018})}\BibitemShut {NoStop}%
\bibitem [{\citenamefont {Bilous}\ \emph {et~al.}(2025)\citenamefont {Bilous}, \citenamefont {Thirion}, \citenamefont {Menke}, \citenamefont {Haverkort}, \citenamefont {P\'alffy},\ and\ \citenamefont {Hansmann}}]{PhysRevB.111.035124}%
  \BibitemOpen
  \bibfield  {author} {\bibinfo {author} {\bibfnamefont {P.}~\bibnamefont {Bilous}}, \bibinfo {author} {\bibfnamefont {L.}~\bibnamefont {Thirion}}, \bibinfo {author} {\bibfnamefont {H.}~\bibnamefont {Menke}}, \bibinfo {author} {\bibfnamefont {M.~W.}\ \bibnamefont {Haverkort}}, \bibinfo {author} {\bibfnamefont {A.}~\bibnamefont {P\'alffy}},\ and\ \bibinfo {author} {\bibfnamefont {P.}~\bibnamefont {Hansmann}},\ }\bibfield  {title} {\bibinfo {title} {Neural-network-supported basis optimizer for the configuration interaction problem in quantum many-body clusters: {Feasibility} study and numerical proof},\ }\href {https://doi.org/10.1103/PhysRevB.111.035124} {\bibfield  {journal} {\bibinfo  {journal} {Phys. Rev. B}\ }\textbf {\bibinfo {volume} {111}},\ \bibinfo {pages} {035124} (\bibinfo {year} {2025})}\BibitemShut {NoStop}%
\bibitem [{\citenamefont {Bilous}\ \emph {et~al.}(2023)\citenamefont {Bilous}, \citenamefont {P\'alffy},\ and\ \citenamefont {Marquardt}}]{PhysRevLett.131.133002}%
  \BibitemOpen
  \bibfield  {author} {\bibinfo {author} {\bibfnamefont {P.}~\bibnamefont {Bilous}}, \bibinfo {author} {\bibfnamefont {A.}~\bibnamefont {P\'alffy}},\ and\ \bibinfo {author} {\bibfnamefont {F.}~\bibnamefont {Marquardt}},\ }\bibfield  {title} {\bibinfo {title} {Deep-learning approach for the atomic configuration interaction problem on large basis sets},\ }\href {https://doi.org/10.1103/PhysRevLett.131.133002} {\bibfield  {journal} {\bibinfo  {journal} {Phys. Rev. Lett.}\ }\textbf {\bibinfo {volume} {131}},\ \bibinfo {pages} {133002} (\bibinfo {year} {2023})}\BibitemShut {NoStop}%
\bibitem [{\citenamefont {Rano}\ and\ \citenamefont {Ghosh}(2023)}]{Rano2023}%
  \BibitemOpen
  \bibfield  {author} {\bibinfo {author} {\bibfnamefont {M.}~\bibnamefont {Rano}}\ and\ \bibinfo {author} {\bibfnamefont {D.}~\bibnamefont {Ghosh}},\ }\bibfield  {title} {\bibinfo {title} {Efficient machine learning configuration interaction for bond breaking problems},\ }\href {https://doi.org/10.1021/acs.jpca.2c09103} {\bibfield  {journal} {\bibinfo  {journal} {The Journal of Physical Chemistry A}\ }\textbf {\bibinfo {volume} {127}},\ \bibinfo {pages} {3705} (\bibinfo {year} {2023})}\BibitemShut {NoStop}%
\bibitem [{\citenamefont {Yu}\ \emph {et~al.}(2025)\citenamefont {Yu}, \citenamefont {Moreno}, \citenamefont {Iosue}, \citenamefont {Bertels}, \citenamefont {Claudino}, \citenamefont {Fuller}, \citenamefont {Groszkowski}, \citenamefont {Humble}, \citenamefont {Jurcevic}, \citenamefont {Kirby}, \citenamefont {Maier}, \citenamefont {Motta}, \citenamefont {Pokharel}, \citenamefont {Seif}, \citenamefont {Shehata}, \citenamefont {Sung}, \citenamefont {Tran}, \citenamefont {Tripathi}, \citenamefont {Mezzacapo},\ and\ \citenamefont {Sharma}}]{yu2025quantumcentricalgorithmsamplebasedkrylov}%
  \BibitemOpen
  \bibfield  {author} {\bibinfo {author} {\bibfnamefont {J.}~\bibnamefont {Yu}}, \bibinfo {author} {\bibfnamefont {J.~R.}\ \bibnamefont {Moreno}}, \bibinfo {author} {\bibfnamefont {J.~T.}\ \bibnamefont {Iosue}}, \bibinfo {author} {\bibfnamefont {L.}~\bibnamefont {Bertels}}, \bibinfo {author} {\bibfnamefont {D.}~\bibnamefont {Claudino}}, \bibinfo {author} {\bibfnamefont {B.}~\bibnamefont {Fuller}}, \bibinfo {author} {\bibfnamefont {P.}~\bibnamefont {Groszkowski}}, \bibinfo {author} {\bibfnamefont {T.~S.}\ \bibnamefont {Humble}}, \bibinfo {author} {\bibfnamefont {P.}~\bibnamefont {Jurcevic}}, \bibinfo {author} {\bibfnamefont {W.}~\bibnamefont {Kirby}}, \bibinfo {author} {\bibfnamefont {T.~A.}\ \bibnamefont {Maier}}, \bibinfo {author} {\bibfnamefont {M.}~\bibnamefont {Motta}}, \bibinfo {author} {\bibfnamefont {B.}~\bibnamefont {Pokharel}}, \bibinfo {author} {\bibfnamefont {A.}~\bibnamefont {Seif}}, \bibinfo {author} {\bibfnamefont {A.}~\bibnamefont {Shehata}}, \bibinfo {author} {\bibfnamefont {K.~J.}\
  \bibnamefont {Sung}}, \bibinfo {author} {\bibfnamefont {M.~C.}\ \bibnamefont {Tran}}, \bibinfo {author} {\bibfnamefont {V.}~\bibnamefont {Tripathi}}, \bibinfo {author} {\bibfnamefont {A.}~\bibnamefont {Mezzacapo}},\ and\ \bibinfo {author} {\bibfnamefont {K.}~\bibnamefont {Sharma}},\ }\href {https://arxiv.org/abs/2501.09702} {\bibinfo {title} {Quantum-centric algorithm for sample-based {Krylov} diagonalization}} (\bibinfo {year} {2025}),\ \Eprint {https://arxiv.org/abs/2501.09702} {arXiv:2501.09702 [quant-ph]} \BibitemShut {NoStop}%
\bibitem [{\citenamefont {Carleo}\ \emph {et~al.}(2019{\natexlab{b}})\citenamefont {Carleo}, \citenamefont {Choo}, \citenamefont {Hofmann}, \citenamefont {Smith}, \citenamefont {Westerhout}, \citenamefont {Alet}, \citenamefont {Davis}, \citenamefont {Efthymiou}, \citenamefont {Glasser}, \citenamefont {Lin}, \citenamefont {Mauri}, \citenamefont {Mazzola}, \citenamefont {Mendl}, \citenamefont {van Nieuwenburg}, \citenamefont {O'Reilly}, \citenamefont {Th{\'e}veniaut}, \citenamefont {Torlai}, \citenamefont {Vicentini},\ and\ \citenamefont {Wietek}}]{netket2:2019}%
  \BibitemOpen
  \bibfield  {author} {\bibinfo {author} {\bibfnamefont {G.}~\bibnamefont {Carleo}}, \bibinfo {author} {\bibfnamefont {K.}~\bibnamefont {Choo}}, \bibinfo {author} {\bibfnamefont {D.}~\bibnamefont {Hofmann}}, \bibinfo {author} {\bibfnamefont {J.~E.~T.}\ \bibnamefont {Smith}}, \bibinfo {author} {\bibfnamefont {T.}~\bibnamefont {Westerhout}}, \bibinfo {author} {\bibfnamefont {F.}~\bibnamefont {Alet}}, \bibinfo {author} {\bibfnamefont {E.~J.}\ \bibnamefont {Davis}}, \bibinfo {author} {\bibfnamefont {S.}~\bibnamefont {Efthymiou}}, \bibinfo {author} {\bibfnamefont {I.}~\bibnamefont {Glasser}}, \bibinfo {author} {\bibfnamefont {S.-H.}\ \bibnamefont {Lin}}, \bibinfo {author} {\bibfnamefont {M.}~\bibnamefont {Mauri}}, \bibinfo {author} {\bibfnamefont {G.}~\bibnamefont {Mazzola}}, \bibinfo {author} {\bibfnamefont {C.~B.}\ \bibnamefont {Mendl}}, \bibinfo {author} {\bibfnamefont {E.}~\bibnamefont {van Nieuwenburg}}, \bibinfo {author} {\bibfnamefont {O.}~\bibnamefont {O'Reilly}}, \bibinfo {author} {\bibfnamefont
  {H.}~\bibnamefont {Th{\'e}veniaut}}, \bibinfo {author} {\bibfnamefont {G.}~\bibnamefont {Torlai}}, \bibinfo {author} {\bibfnamefont {F.}~\bibnamefont {Vicentini}},\ and\ \bibinfo {author} {\bibfnamefont {A.}~\bibnamefont {Wietek}},\ }\bibfield  {title} {\bibinfo {title} {Netket: {A} machine learning toolkit for many-body quantum systems},\ }\href {https://doi.org/10.1016/j.softx.2019.100311} {\bibfield  {journal} {\bibinfo  {journal} {SoftwareX}\ ,\ \bibinfo {pages} {100311}} (\bibinfo {year} {2019}{\natexlab{b}})}\BibitemShut {NoStop}%
\bibitem [{\citenamefont {Vicentini}\ \emph {et~al.}(2022)\citenamefont {Vicentini}, \citenamefont {Hofmann}, \citenamefont {Szabó}, \citenamefont {Wu}, \citenamefont {Roth}, \citenamefont {Giuliani}, \citenamefont {Pescia}, \citenamefont {Nys}, \citenamefont {Vargas-Calderón}, \citenamefont {Astrakhantsev},\ and\ \citenamefont {Carleo}}]{netket3:2022}%
  \BibitemOpen
  \bibfield  {author} {\bibinfo {author} {\bibfnamefont {F.}~\bibnamefont {Vicentini}}, \bibinfo {author} {\bibfnamefont {D.}~\bibnamefont {Hofmann}}, \bibinfo {author} {\bibfnamefont {A.}~\bibnamefont {Szabó}}, \bibinfo {author} {\bibfnamefont {D.}~\bibnamefont {Wu}}, \bibinfo {author} {\bibfnamefont {C.}~\bibnamefont {Roth}}, \bibinfo {author} {\bibfnamefont {C.}~\bibnamefont {Giuliani}}, \bibinfo {author} {\bibfnamefont {G.}~\bibnamefont {Pescia}}, \bibinfo {author} {\bibfnamefont {J.}~\bibnamefont {Nys}}, \bibinfo {author} {\bibfnamefont {V.}~\bibnamefont {Vargas-Calderón}}, \bibinfo {author} {\bibfnamefont {N.}~\bibnamefont {Astrakhantsev}},\ and\ \bibinfo {author} {\bibfnamefont {G.}~\bibnamefont {Carleo}},\ }\bibfield  {title} {\bibinfo {title} {{NetKet 3: Machine Learning Toolbox for Many-Body Quantum Systems}},\ }\href {https://doi.org/10.21468/SciPostPhysCodeb.7} {\bibfield  {journal} {\bibinfo  {journal} {SciPost Phys. Codebases}\ ,\ \bibinfo {pages} {7}} (\bibinfo {year} {2022})}\BibitemShut
  {NoStop}%
\bibitem [{\citenamefont {Pfeuty}(1970)}]{PFEUTY197079}%
  \BibitemOpen
  \bibfield  {author} {\bibinfo {author} {\bibfnamefont {P.}~\bibnamefont {Pfeuty}},\ }\bibfield  {title} {\bibinfo {title} {The one-dimensional {Ising} model with a transverse field},\ }\href {https://doi.org/https://doi.org/10.1016/0003-4916(70)90270-8} {\bibfield  {journal} {\bibinfo  {journal} {Annals of Physics}\ }\textbf {\bibinfo {volume} {57}},\ \bibinfo {pages} {79} (\bibinfo {year} {1970})}\BibitemShut {NoStop}%
\bibitem [{\citenamefont {Young}\ and\ \citenamefont {Rieger}(1996)}]{PhysRevB.53.8486}%
  \BibitemOpen
  \bibfield  {author} {\bibinfo {author} {\bibfnamefont {A.~P.}\ \bibnamefont {Young}}\ and\ \bibinfo {author} {\bibfnamefont {H.}~\bibnamefont {Rieger}},\ }\bibfield  {title} {\bibinfo {title} {Numerical study of the random transverse-field {Ising} spin chain},\ }\href {https://doi.org/10.1103/PhysRevB.53.8486} {\bibfield  {journal} {\bibinfo  {journal} {Phys. Rev. B}\ }\textbf {\bibinfo {volume} {53}},\ \bibinfo {pages} {8486} (\bibinfo {year} {1996})}\BibitemShut {NoStop}%
\bibitem [{\citenamefont {Becca}\ and\ \citenamefont {Sorella}(2017)}]{becca2017quantum}%
  \BibitemOpen
  \bibfield  {author} {\bibinfo {author} {\bibfnamefont {F.}~\bibnamefont {Becca}}\ and\ \bibinfo {author} {\bibfnamefont {S.}~\bibnamefont {Sorella}},\ }\href@noop {} {\emph {\bibinfo {title} {Quantum {Monte Carlo} approaches for correlated systems}}}\ (\bibinfo  {publisher} {Cambridge University Press},\ \bibinfo {year} {2017})\BibitemShut {NoStop}%
\bibitem [{\citenamefont {Brodoloni}\ and\ \citenamefont {Pilati}(2024)}]{PhysRevE.110.065305}%
  \BibitemOpen
  \bibfield  {author} {\bibinfo {author} {\bibfnamefont {L.}~\bibnamefont {Brodoloni}}\ and\ \bibinfo {author} {\bibfnamefont {S.}~\bibnamefont {Pilati}},\ }\bibfield  {title} {\bibinfo {title} {Zero-temperature {Monte Carlo} simulations of two-dimensional quantum spin glasses guided by neural network states},\ }\href {https://doi.org/10.1103/PhysRevE.110.065305} {\bibfield  {journal} {\bibinfo  {journal} {Phys. Rev. E}\ }\textbf {\bibinfo {volume} {110}},\ \bibinfo {pages} {065305} (\bibinfo {year} {2024})}\BibitemShut {NoStop}%
\bibitem [{\citenamefont {Park}\ and\ \citenamefont {Kastoryano}(2020)}]{PhysRevResearch.2.023232}%
  \BibitemOpen
  \bibfield  {author} {\bibinfo {author} {\bibfnamefont {C.-Y.}\ \bibnamefont {Park}}\ and\ \bibinfo {author} {\bibfnamefont {M.~J.}\ \bibnamefont {Kastoryano}},\ }\bibfield  {title} {\bibinfo {title} {Geometry of learning neural quantum states},\ }\href {https://doi.org/10.1103/PhysRevResearch.2.023232} {\bibfield  {journal} {\bibinfo  {journal} {Phys. Rev. Res.}\ }\textbf {\bibinfo {volume} {2}},\ \bibinfo {pages} {023232} (\bibinfo {year} {2020})}\BibitemShut {NoStop}%
\bibitem [{\citenamefont {Reinholdt}\ \emph {et~al.}(2025)\citenamefont {Reinholdt}, \citenamefont {Ziems}, \citenamefont {Kjellgren}, \citenamefont {Coriani}, \citenamefont {Sauer},\ and\ \citenamefont {Kongsted}}]{reinholdt2025fundamentallimitationssamplebasedquantum}%
  \BibitemOpen
  \bibfield  {author} {\bibinfo {author} {\bibfnamefont {P.}~\bibnamefont {Reinholdt}}, \bibinfo {author} {\bibfnamefont {K.~M.}\ \bibnamefont {Ziems}}, \bibinfo {author} {\bibfnamefont {E.~R.}\ \bibnamefont {Kjellgren}}, \bibinfo {author} {\bibfnamefont {S.}~\bibnamefont {Coriani}}, \bibinfo {author} {\bibfnamefont {S.~P.~A.}\ \bibnamefont {Sauer}},\ and\ \bibinfo {author} {\bibfnamefont {J.}~\bibnamefont {Kongsted}},\ }\href {https://arxiv.org/abs/2501.07231} {\bibinfo {title} {Fundamental limitations in sample-based quantum diagonalization methods}} (\bibinfo {year} {2025}),\ \Eprint {https://arxiv.org/abs/2501.07231} {arXiv:2501.07231 [physics.chem-ph]} \BibitemShut {NoStop}%
\bibitem [{\citenamefont {Ho}\ and\ \citenamefont {Hsieh}(2019)}]{Ho_2019}%
  \BibitemOpen
  \bibfield  {author} {\bibinfo {author} {\bibfnamefont {W.~W.}\ \bibnamefont {Ho}}\ and\ \bibinfo {author} {\bibfnamefont {T.~H.}\ \bibnamefont {Hsieh}},\ }\bibfield  {title} {\bibinfo {title} {Efficient variational simulation of non-trivial quantum states},\ }\bibfield  {journal} {\bibinfo  {journal} {SciPost Physics}\ }\textbf {\bibinfo {volume} {6}},\ \href {https://doi.org/10.21468/scipostphys.6.3.029} {10.21468/scipostphys.6.3.029} (\bibinfo {year} {2019})\BibitemShut {NoStop}%
\bibitem [{\citenamefont {Asthana}\ \emph {et~al.}(2023)\citenamefont {Asthana}, \citenamefont {Kumar}, \citenamefont {Abraham}, \citenamefont {Grimsley}, \citenamefont {Zhang}, \citenamefont {Cincio}, \citenamefont {Tretiak}, \citenamefont {Dub}, \citenamefont {Economou}, \citenamefont {Barnes},\ and\ \citenamefont {Mayhall}}]{D2SC05371C}%
  \BibitemOpen
  \bibfield  {author} {\bibinfo {author} {\bibfnamefont {A.}~\bibnamefont {Asthana}}, \bibinfo {author} {\bibfnamefont {A.}~\bibnamefont {Kumar}}, \bibinfo {author} {\bibfnamefont {V.}~\bibnamefont {Abraham}}, \bibinfo {author} {\bibfnamefont {H.}~\bibnamefont {Grimsley}}, \bibinfo {author} {\bibfnamefont {Y.}~\bibnamefont {Zhang}}, \bibinfo {author} {\bibfnamefont {L.}~\bibnamefont {Cincio}}, \bibinfo {author} {\bibfnamefont {S.}~\bibnamefont {Tretiak}}, \bibinfo {author} {\bibfnamefont {P.~A.}\ \bibnamefont {Dub}}, \bibinfo {author} {\bibfnamefont {S.~E.}\ \bibnamefont {Economou}}, \bibinfo {author} {\bibfnamefont {E.}~\bibnamefont {Barnes}},\ and\ \bibinfo {author} {\bibfnamefont {N.~J.}\ \bibnamefont {Mayhall}},\ }\bibfield  {title} {\bibinfo {title} {Quantum self-consistent equation-of-motion method for computing molecular excitation energies{,} ionization potentials{,} and electron affinities on a quantum computer},\ }\href {https://doi.org/10.1039/D2SC05371C} {\bibfield  {journal} {\bibinfo  {journal}
  {Chem. Sci.}\ }\textbf {\bibinfo {volume} {14}},\ \bibinfo {pages} {2405} (\bibinfo {year} {2023})}\BibitemShut {NoStop}%
\bibitem [{\citenamefont {Cantori}()}]{Cantori_Adaptive-basis_sample-based_neural}%
  \BibitemOpen
  \bibfield  {author} {\bibinfo {author} {\bibfnamefont {S.}~\bibnamefont {Cantori}},\ }\href {https://github.com/simonecantori/Sample-based-Neural-Diagonalization} {\bibinfo {title} {{Adaptive-basis sample-based neural diagonalization for quantum many-body systems}}},\ \bibinfo {howpublished} {GitHub repository}\BibitemShut {NoStop}%
\bibitem [{\citenamefont {Li}\ \emph {et~al.}(2017)\citenamefont {Li}, \citenamefont {Yang}, \citenamefont {Peng},\ and\ \citenamefont {Sun}}]{PhysRevLett.118.150503}%
  \BibitemOpen
  \bibfield  {author} {\bibinfo {author} {\bibfnamefont {J.}~\bibnamefont {Li}}, \bibinfo {author} {\bibfnamefont {X.}~\bibnamefont {Yang}}, \bibinfo {author} {\bibfnamefont {X.}~\bibnamefont {Peng}},\ and\ \bibinfo {author} {\bibfnamefont {C.-P.}\ \bibnamefont {Sun}},\ }\bibfield  {title} {\bibinfo {title} {Hybrid quantum-classical approach to quantum optimal control},\ }\href {https://doi.org/10.1103/PhysRevLett.118.150503} {\bibfield  {journal} {\bibinfo  {journal} {Phys. Rev. Lett.}\ }\textbf {\bibinfo {volume} {118}},\ \bibinfo {pages} {150503} (\bibinfo {year} {2017})}\BibitemShut {NoStop}%
\bibitem [{\citenamefont {Mitarai}\ \emph {et~al.}(2018)\citenamefont {Mitarai}, \citenamefont {Negoro}, \citenamefont {Kitagawa},\ and\ \citenamefont {Fujii}}]{PhysRevA.98.032309}%
  \BibitemOpen
  \bibfield  {author} {\bibinfo {author} {\bibfnamefont {K.}~\bibnamefont {Mitarai}}, \bibinfo {author} {\bibfnamefont {M.}~\bibnamefont {Negoro}}, \bibinfo {author} {\bibfnamefont {M.}~\bibnamefont {Kitagawa}},\ and\ \bibinfo {author} {\bibfnamefont {K.}~\bibnamefont {Fujii}},\ }\bibfield  {title} {\bibinfo {title} {Quantum circuit learning},\ }\href {https://doi.org/10.1103/PhysRevA.98.032309} {\bibfield  {journal} {\bibinfo  {journal} {Phys. Rev. A}\ }\textbf {\bibinfo {volume} {98}},\ \bibinfo {pages} {032309} (\bibinfo {year} {2018})}\BibitemShut {NoStop}%
\bibitem [{\citenamefont {Paszke}\ \emph {et~al.}(2017)\citenamefont {Paszke}, \citenamefont {Gross}, \citenamefont {Chintala}, \citenamefont {Chanan}, \citenamefont {Yang}, \citenamefont {DeVito}, \citenamefont {Lin}, \citenamefont {Desmaison}, \citenamefont {Antiga},\ and\ \citenamefont {Lerer}}]{paszke2017automatic}%
  \BibitemOpen
  \bibfield  {author} {\bibinfo {author} {\bibfnamefont {A.}~\bibnamefont {Paszke}}, \bibinfo {author} {\bibfnamefont {S.}~\bibnamefont {Gross}}, \bibinfo {author} {\bibfnamefont {S.}~\bibnamefont {Chintala}}, \bibinfo {author} {\bibfnamefont {G.}~\bibnamefont {Chanan}}, \bibinfo {author} {\bibfnamefont {E.}~\bibnamefont {Yang}}, \bibinfo {author} {\bibfnamefont {Z.}~\bibnamefont {DeVito}}, \bibinfo {author} {\bibfnamefont {Z.}~\bibnamefont {Lin}}, \bibinfo {author} {\bibfnamefont {A.}~\bibnamefont {Desmaison}}, \bibinfo {author} {\bibfnamefont {L.}~\bibnamefont {Antiga}},\ and\ \bibinfo {author} {\bibfnamefont {A.}~\bibnamefont {Lerer}},\ }\bibfield  {title} {\bibinfo {title} {Automatic differentiation in pytorch},\ }in\ \href@noop {} {\emph {\bibinfo {booktitle} {NIPS-W}}}\ (\bibinfo {year} {2017})\BibitemShut {NoStop}%
\bibitem [{\citenamefont {Mardia}\ and\ \citenamefont {Jupp}(2009)}]{mardia2009directional}%
  \BibitemOpen
  \bibfield  {author} {\bibinfo {author} {\bibfnamefont {K.}~\bibnamefont {Mardia}}\ and\ \bibinfo {author} {\bibfnamefont {P.}~\bibnamefont {Jupp}},\ }\href {https://books.google.it/books?id=PTNiCm4Q-M0C} {\emph {\bibinfo {title} {Directional Statistics}}},\ Wiley Series in Probability and Statistics\ (\bibinfo  {publisher} {Wiley},\ \bibinfo {year} {2009})\BibitemShut {NoStop}%
\bibitem [{\citenamefont {Xuan}\ \emph {et~al.}(2025)\citenamefont {Xuan}, \citenamefont {Yang},\ and\ \citenamefont {Li}}]{xuan2025exploringimpacttemperaturescaling}%
  \BibitemOpen
  \bibfield  {author} {\bibinfo {author} {\bibfnamefont {H.}~\bibnamefont {Xuan}}, \bibinfo {author} {\bibfnamefont {B.}~\bibnamefont {Yang}},\ and\ \bibinfo {author} {\bibfnamefont {X.}~\bibnamefont {Li}},\ }\href {https://arxiv.org/abs/2502.20604} {\bibinfo {title} {Exploring the impact of temperature scaling in softmax for classification and adversarial robustness}} (\bibinfo {year} {2025}),\ \Eprint {https://arxiv.org/abs/2502.20604} {arXiv:2502.20604 [cs.LG]} \BibitemShut {NoStop}%
\bibitem [{\citenamefont {Vaswani}\ \emph {et~al.}(2023)\citenamefont {Vaswani}, \citenamefont {Shazeer}, \citenamefont {Parmar}, \citenamefont {Uszkoreit}, \citenamefont {Jones}, \citenamefont {Gomez}, \citenamefont {Kaiser},\ and\ \citenamefont {Polosukhin}}]{vaswani2023attentionneed}%
  \BibitemOpen
  \bibfield  {author} {\bibinfo {author} {\bibfnamefont {A.}~\bibnamefont {Vaswani}}, \bibinfo {author} {\bibfnamefont {N.}~\bibnamefont {Shazeer}}, \bibinfo {author} {\bibfnamefont {N.}~\bibnamefont {Parmar}}, \bibinfo {author} {\bibfnamefont {J.}~\bibnamefont {Uszkoreit}}, \bibinfo {author} {\bibfnamefont {L.}~\bibnamefont {Jones}}, \bibinfo {author} {\bibfnamefont {A.~N.}\ \bibnamefont {Gomez}}, \bibinfo {author} {\bibfnamefont {L.}~\bibnamefont {Kaiser}},\ and\ \bibinfo {author} {\bibfnamefont {I.}~\bibnamefont {Polosukhin}},\ }\href {https://arxiv.org/abs/1706.03762} {\bibinfo {title} {Attention is all you need}} (\bibinfo {year} {2023}),\ \Eprint {https://arxiv.org/abs/1706.03762} {arXiv:1706.03762 [cs.CL]} \BibitemShut {NoStop}%
\end{thebibliography}%

\appendix 
%\section{} \label{app}

\section{Loss function of SND and its gradient}\label{app1}
As in Eq.~\eqref{loss}, one can write the probability of a set of configurations $S^{(k)}$ as $P(S^{(k)})=\prod_{x^{(l)}\in S^{(k)}}P(x^{(l)})$, where $P(x^{(l)})$ is the probability of sampling the bitstring $x^{(l)}$ defined by an autoregressive NN.
The loss function $L$ and its derivative with respect to a parameter of the network $\omega$ can be calculated as follows:
\begin{equation}
 \begin{aligned}
    L&=\sum_k \biggl(\prod_{x^{(l)}\in S^{(k)}}P(x^{(l)})\biggr)E^{(k)}\Longrightarrow \\
    \frac{\partial L}{\partial\omega} &= \sum_k \frac{\partial}{\partial\omega}\biggl(\prod_{x^{(l)}\in S^{(k)}}P(x^{(l)})\biggr)E^{(k)} =
    \\
    & = \sum_k \biggl[\sum_{x^{(l)}\in S^{(k)}}\biggl(\frac{\partial P(x^{(l)})}{\partial\omega}\prod_{x^{(m)}\neq x^{(l)}}P(x^{(m)})\biggr)\biggr]E^{(k)} \, ,
 \end{aligned}
\end{equation}
where $x^{(l)}$ and $x^{(m)}$ are bitstrings in the batch $S^{(k)}$. 
With a straightforward rearrangement, the derivative can be rewritten as
%$\frac{P(x^{(l)})}{P(x^{(l)})}=1$:
%
\begin{equation}
 \begin{aligned}
    \frac{\partial L}{\partial\omega} & = \sum_k \biggl[\sum_{x^{(l)}\in S^{(k)}}\biggl(\frac{\partial P(x^{(l)})}{\partial\omega}\frac{P(x^{(l)})}{P(x^{(l)})}\prod_{x^{(m)}\neq x^{(l)}}P(x^{(m)})\biggr)\biggr]E^{(k)} 
    \\
    & = \sum_k P(S^{(k)}) \biggl( \sum_{x^{(l)}\in S^{(k)}} \frac{\partial \log(P(x^{(l)}))}{\partial\omega} \biggr ) E^{(k)} \, ,
 \end{aligned}
\end{equation}
and the stochastic estimator is given by
\begin{equation}
    \frac{\partial L}{\partial\omega} \simeq \frac{1}{K}\sum_{k=1}^K\biggl( \sum_{x^{(l)}\in S^{(k)}} \frac{\partial \log(P(x^{(l)}))}{\partial\omega} \biggr ) E^{(k)} \, ,
\end{equation}
where the batches of bitstrings $S^{(k)}$ are sampled according to $P(S^{(k)})$.
A baseline term is useful to stabilize the training process~\cite{Barrett2022,PhysRevLett.124.020503}. In our framework, we set it equal to the average energy over the $K$ batches $\overline{E^{(k)}}$. Therefore, the loss function for SND reads:
\begin{equation}
  L=\frac{1}{K} \sum_{k=1}^{K}\biggl(\sum_{x^{(l)}\in S^{(k)}}\log(P(x^{(l)}))\biggr)(E^{(k)}-\overline{E^{(k)}}) \, .
\end{equation}

\section{Loss function of AB-SND and its gradient}\label{app2}
For the AB-SND method, the derivative with respect to the parameters of the NN used to sample basis configurations $\frac{\partial L}{\partial\omega}$ is calculated as discussed in the previous section. However, here we also want to optimize the basis-change parameters $\boldsymbol{\theta}$ in order to minimize the estimated ground-state energy. Notably, the rotation angles $\boldsymbol{\theta}$ are also used as a condition for the NN that generates the spin configurations. For this, they are provided as inputs preceding the spin values.
Similarly to the SND method, one obtains:
\begin{equation}
\begin{aligned}
    \frac{\partial L}{\partial\theta_i} = &\sum_k \frac{\partial}{\partial\theta_i}\biggl(\prod_{x^{(l)}\in S^{(k)}}P(x^{(l)}|\boldsymbol{\theta})\biggr)E^{(k)} \\ &+ \biggl(\prod_{x^{(l)}\in S^{(k)}}P(x^{(l)}|\boldsymbol{\theta})\biggr)\frac{\partial}{\partial\theta_i}E^{(k)} \, .
\end{aligned}
\end{equation}
From the Hellmann–Feynman theorem, one can write
\begin{equation}
    \frac{\partial E^{(k)}(\theta_i)}{\partial \theta_i} = \bigg\langle\psi_0^{(k)}\bigg|\frac{\partial\hat{H}}{\partial \theta_i}\bigg|\psi_0^{(k)}\bigg\rangle \, ,
\end{equation}
where $\psi_0^{(k)}$ is the estimated ground-state wave function.
In AB-SND, we use a basis-change unitary operator $U(\boldsymbol{\theta})$, so
\begin{equation}
    \frac{\partial E^{(k)}(\theta_i)}{\partial \theta_i} = \bigg\langle\psi_0^{(k)}\bigg|\frac{\partial(U^\dag (\boldsymbol{\theta})\hat{H} U(\boldsymbol{\theta}))}{\partial \theta_i}\bigg|\psi_0^{(k)}\bigg\rangle \, .
\end{equation}
This quantity can be calculated using the parameter-shift rule~\cite{PhysRevLett.118.150503, PhysRevA.98.032309}. Alternatively, if $U(\boldsymbol{\theta})$ is implemented using classical algorithms, the gradient can be calculated using automatic differentiation, e.g., via the Pytorch library~\cite{paszke2017automatic}.
Finally, the derivative with respect to a generic rotation angle reads:
\begin{equation}
\begin{aligned}
    \frac{\partial L}{\partial\theta_i} \simeq & \frac{1}{K}\sum_{k=1}^K\bigg[\biggl( \sum_{x^{(l)}\in S^{(k)}} \frac{\partial \log(P(x^{(l)}|\boldsymbol{\theta}))}{\partial\theta_i} \biggr ) E^{(k)} \\& + \bigg\langle\psi_0^{(k)}\bigg|\frac{\partial(U^\dag (\boldsymbol{\theta})\hat{H} U(\boldsymbol{\theta}))}{\partial \theta_i}\bigg|\psi_0^{(k)}\bigg\rangle \biggl]  \, .
\end{aligned}
\end{equation}
We also implement an alternative approach to optimize the rotation angles, which avoids the multiple diagonalization steps used in the parameter-shift rule. This approach involves sampling the angles $\boldsymbol{\theta}$ from an additional autoregressive NN. It is detailed in Appendix~\ref{app5}.

\section{Failure of standard SBD approaches at large transverse field}\label{app3}
\begin{figure*}
	\centering
	\includegraphics[width=\textwidth]{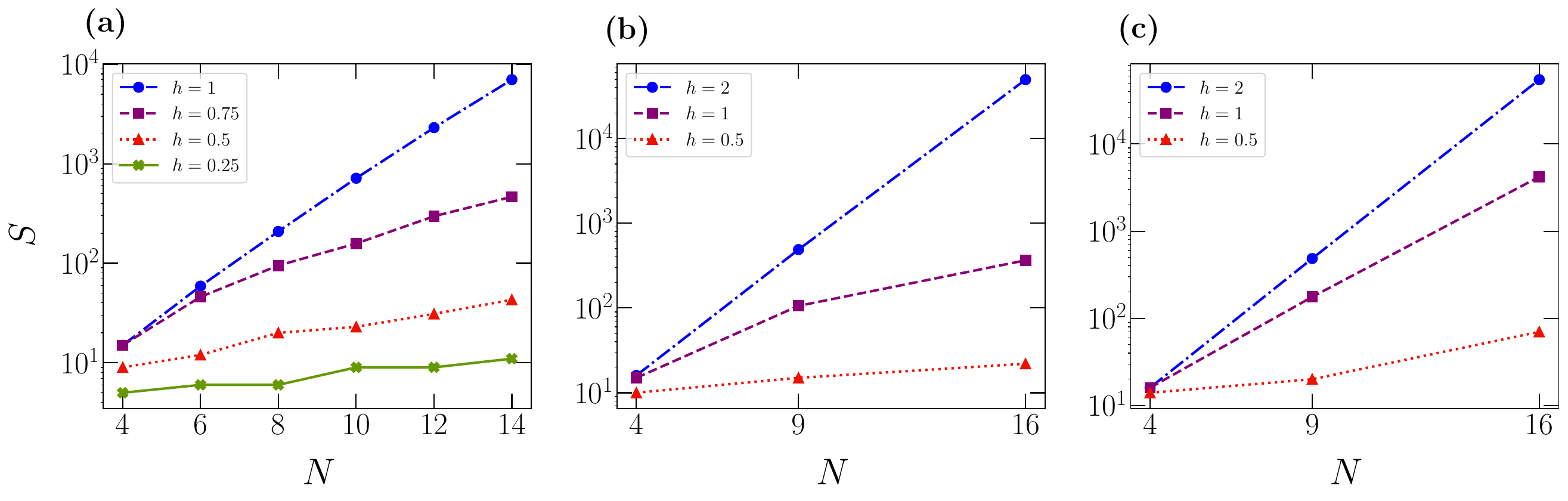}
	\caption{Number of unique configurations $S$ required to reach the relative error $\epsilon=0.01$ as a function of number of spins $N$ using a standard SBD approach with configurations sampled from the exact ground state. Panels (a), (b), and (c) display results for the 1D-TFIM, the 2D-TFIM, and the 2D-EAM, respectively. Different datasets in each panel correspond to different transverse fields $h$.} 
	\label{gs}
\end{figure*}

The accuracy of SBD approaches noticeably depends on how the computational basis elements used to build the truncated Hamiltonian matrix are sampled. In Fig.~\ref{gs}, we show that even when the exact ground-state wave function is used for sampling, without adaptive basis rotations the SBD method fails when the ground-state wave function is not strongly concentrated in the chosen computational basis. In fact, beyond the small transverse field regime $h \ll 1$, the truncated basis size $S$ required to reach the target accuracy of $1\%$ approaches an exponential scaling with the system size $N$.

\section{Most challenging regime for AB-SND}\label{app4}
\begin{figure}
\centering
\includegraphics[width=\columnwidth]{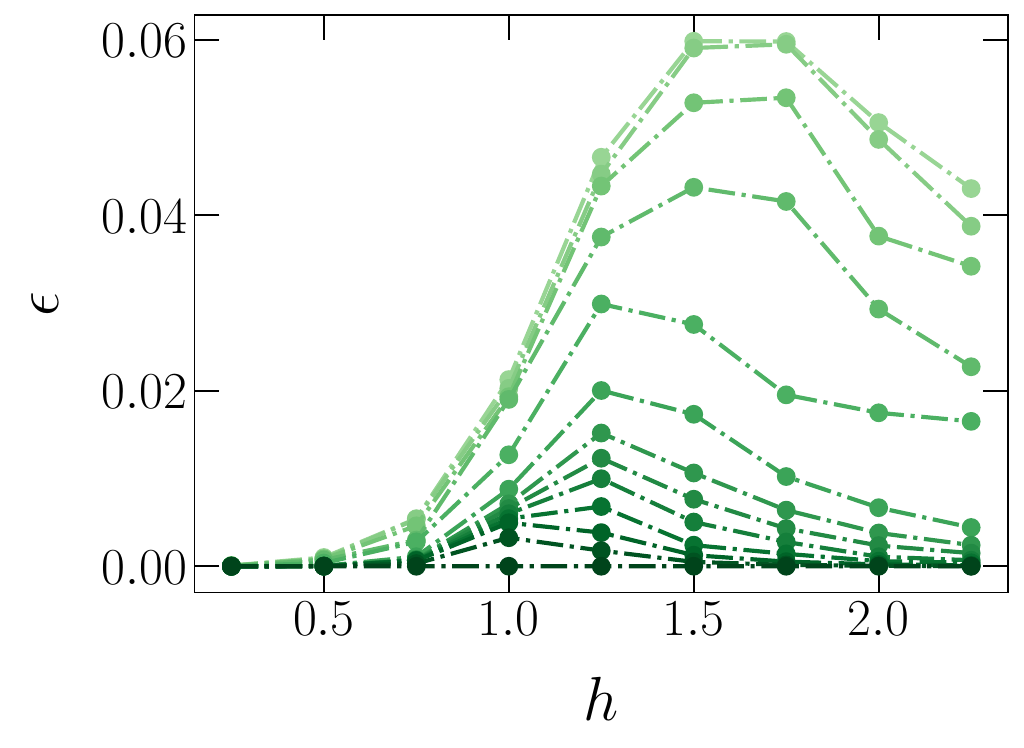}
\caption{Relative error $\epsilon$ as a function of the transverse field $h$, for different numbers of unique configuration $S$. Darker colors mean higher $S$, with $S\in\{4,8,16,32,64,128,256,384,512,640,768,896,1024\}$). The configurations are sampled from the exact ground state. Adaptive single-spin rotations are then applied, as explained in the main text. These results are for the 1D-TFIM with $N=10$ spins.}
\label{h_gs}
\end{figure}

In Fig.~\ref{h_gs} we report numerical evidence showing that, in the large $S$ limit, the peak of the energy error obtained via the AB-SND method drifts towards the critical point of the ferromagnetic quantum phase transition. The chosen testbed is the 1D-TFIM with $N=10$ spins. The configurations are sampled from the exact ground state.
%In Fig.~\ref{h} of the main text, we have shown that for AB-SND methods used for 1D-TFIM, the error for $h=1.5$ is higher than for $h_c=1$, that is the critical point. As explained in the main text, we expect that the most challenging point is at $h_c$. In Fig.~\ref{h_gs}, we show that for large enough values of $S$, the expected behaviour is retrieved. Here, we apply the different single-spin rotations (same for each spin) to the samples obtained from the ideal ground-state. Then, we display the result related to the rotation that gives the best accuracy. Here, the simulation can be done on the ideal ground state because we use $N=10$. 
%

\section{Loss function of AB-SND with sampled rotational parameters}\label{app5}
Instead of optimizing the basis change using the Hellmann–Feynman theorem, one can employ an additional autoregressive NN to sample the parameters $\boldsymbol{\theta}$ that define the basis-change unitary operator. 
%This can be useful if we have degenerate eigenvalues, where standard formulation of Hellmann–Feynman theorem doesn't work anymore (\textcolor{red}{Non sono sicuro di questa cosa, anche la letteratura mi pare un po' combattuta, forse questa intera sezione è inutile alla fine}).
In this alternative procedure, the probability of a set of spin configurations is written as: $P(S^{(k)})=P_{\boldsymbol{\nu}}(\boldsymbol{\theta}^{(k)})P_{\boldsymbol{\omega}}(S^{(k)}|\boldsymbol{\theta}^{(k)})=P_{\boldsymbol{\nu}}(\boldsymbol{\theta}^{(k)})\prod_{x^{(l)}\in S^{(k)}}P_{\boldsymbol{\omega}}(x^{(l)}|\boldsymbol{\theta}^{(k)})$, where $\boldsymbol{\nu}$ denotes the weights of the neural network responsible for sampling the parameters $\boldsymbol{\theta}^{(k)}$, while $\boldsymbol{\omega}$ denotes the weights of the neural network responsible for sampling the bitstrings in the batch $S^{(k)}$. The former, is an autoregressive NN that provides the parameters $\boldsymbol{\mu}$ and $\boldsymbol{\kappa}$ of a Von Mises distribution~\cite{mardia2009directional} from which the basis-change parameters $\boldsymbol{\theta}^{(k)}$ are sampled. Due to the autoregressive architecture, $\mu_i$ and $\kappa_i$ depend on the angles $\theta_j$, with $j<i$. The rotation angles $\boldsymbol{\theta}^{(k)}$ are also used as conditions for the other autoregressive NN that defines the probabilities $P_{\boldsymbol{\omega}}(x^{(l)}|\boldsymbol{\theta}^{(k)})$ of each  bitstrings $x^{(l)}$.
The loss function to be minimized is defined as: 
\begin{equation}
 \begin{aligned}
        L &= \sum_k\int d\boldsymbol{\theta}^{(k)}\biggl[P_{\boldsymbol{\nu}}(\boldsymbol{\theta}^{(k)})\prod_{x^{(l)}\in S^{(k)}}P_{\boldsymbol{\omega}}(x^{(l)}|\boldsymbol{\theta}^{(k)})\biggr]E^{(k)}
 \end{aligned}
\end{equation}
and its derivatives are 
\begin{equation}\label{theta}
 \begin{aligned}
        \frac{\partial L}{\partial\omega} &= \sum_k\int d\boldsymbol{\theta}^{(k)}\biggl[P_{\boldsymbol{\nu}}(\boldsymbol{\theta}^{(k)})\frac{\partial}{\partial\omega}\prod_{x^{(l)}\in S^{(k)}}P_{\boldsymbol{\omega}}(x^{(l)}|\boldsymbol{\theta}^{(k)})\biggr]E^{(k)}
        \\
        & = \sum_k\int d\boldsymbol{\theta}^{(k)}\biggl[P_{\boldsymbol{\nu}}(\boldsymbol{\theta}^{(k)})P_{\boldsymbol{\omega}}(S^{(k)}|\boldsymbol{\theta}^{(k)}) \\
        & \times \biggl( \sum_{x^{(l)}\in S^{(k)}} \frac{\partial \log(P_{\boldsymbol{\omega}}(x^{(l)}|\boldsymbol{\theta}^{(k)}))}{\partial\omega} \biggr )\biggr]E^{(k)}
        %\\
        %& = \sum_k\int d\boldsymbol{\theta}^{(k)}P(S^{(k)})\biggl[\sum_{x^{(l)}\in S^{(k)}} \frac{\partial \log(P_{\boldsymbol{\omega}}(x^{(l)}|\boldsymbol{\theta}^{(k)}))}{\partial\omega}\biggr]E^{(k)}
        \\
        & \simeq \frac{1}{K} \sum_{k=1}^K\biggl[\sum_{x^{(l)}\in S^{(k)}} \frac{\partial \log(P_{\boldsymbol{\omega}}(x^{(l)}|\boldsymbol{\theta}^{(k)}))}{\partial\omega}\biggr]E^{(k)} \, ,
 \end{aligned}
\end{equation}
and
\begin{equation}\label{omega}
 \begin{aligned}
        \frac{\partial L}{\partial\nu} &= \sum_k\int d\boldsymbol{\theta}^{(k)}\biggl[\frac{\partial P_{\boldsymbol{\nu}}(\boldsymbol{\theta}^{(k)})}{\partial\nu}\prod_{x^{(l)}\in S^{(k)}}P_{\boldsymbol{\omega}}(x^{(l)}|\boldsymbol{\theta}^{(k)})\biggr]E^{(k)}
        \\
        & = \sum_k\int d\boldsymbol{\theta}^{(k)}\biggl[\frac{\partial P_{\boldsymbol{\nu}}(\boldsymbol{\theta}^{(k)})}{\partial\nu}\frac{P_{\boldsymbol{\nu}}(\boldsymbol{\theta}^{(k)})}{P_{\boldsymbol{\nu}}(\boldsymbol{\theta}^{(k)})}\prod_{x^{(l)}\in S^{(k)}}P_{\boldsymbol{\omega}}(x^{(l)}|\boldsymbol{\theta}^{(k)})\biggr]E^{(k)}
        \\
        & = \sum_k\int d\boldsymbol{\theta}^{(k)}P_{\boldsymbol{\nu}}(\boldsymbol{\theta}^{(k)})P_{\boldsymbol{\omega}}(S^{(k)}|\boldsymbol{\theta}^{(k)})\biggl[\frac{\partial \log (P_{\boldsymbol{\nu}}(\boldsymbol{\theta}^{(k)}))}{\partial\nu}\biggr]E^{(k)}
        \\
        & \simeq \frac{1}{K} \sum_{k=1}^{K}\biggl[\frac{\partial \log (P_{\boldsymbol{\nu}}(\boldsymbol{\theta}^{(k)}))}{\partial\nu}\biggr]E^{(k)} \, ,
 \end{aligned}
\end{equation}
where $\omega \in \boldsymbol{\omega}$ and $\nu \in \boldsymbol{\nu}$ denote single weights of the corresponding neural networks. %The expressions for the other weights follow analogously.

\begin{figure}
	\centering
	\includegraphics[width=\columnwidth]{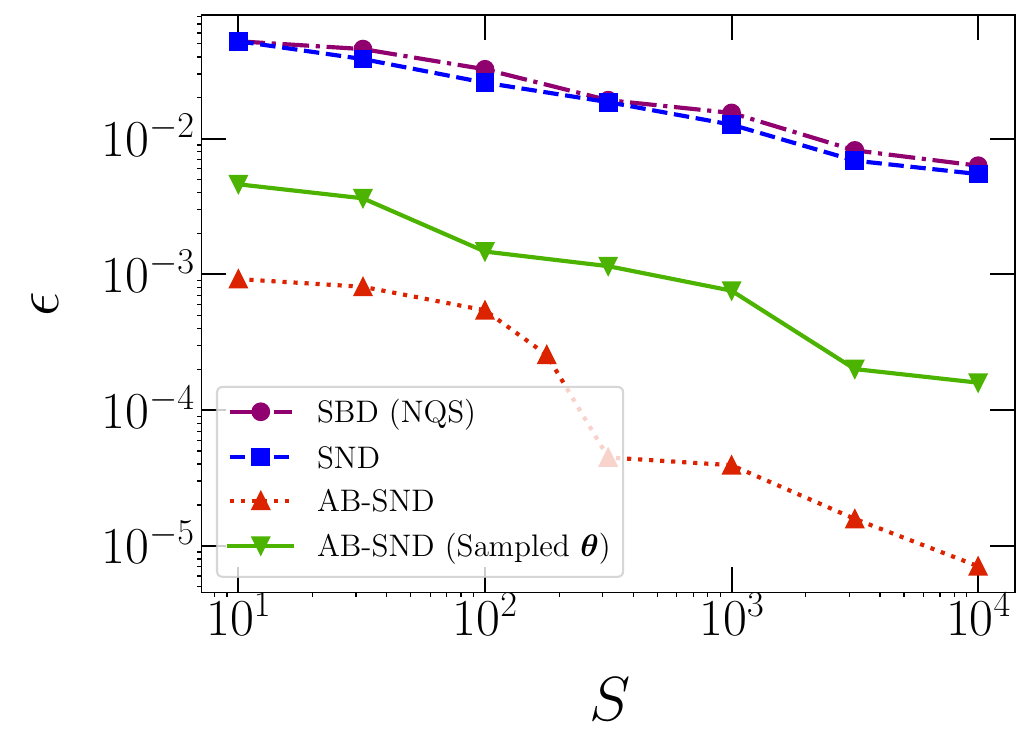}
	\caption{Relative error $\epsilon$ as a function of the number of unique configurations $S$ used to build the subspace Hamiltonian. These results are for the 1D-TFIM with $h=0.5$ and $N=50$. We compare the accuracies of a standard SBD approach powered by NQS sampling, the SND approach, and two AB-SND approaches. The (green) upside-down triangles refer to the AB-SND approach with rotation parameters $\boldsymbol{\theta}^{(k)}$ sampled from an autoregressive NN, as explained in the text.}
	\label{vm_plot}
\end{figure}

%In this way, when we calculate the derivative with respect to $\omega$, we get Eq.\eqref{theta}, while we get Eq.\eqref{omega} for the derivatives with respect to $\nu$.
%
Therefore, including the baseline term, the loss is evaluated as:
\begin{equation}
\begin{aligned}
  L& =\frac{1}{K} \sum_{k=1}^{K}\biggl[\log (P_{\boldsymbol{\nu}}(\boldsymbol{\theta}^{(k)}))+\biggl(\sum_{x^{(l)}\in S^{(k)}}\log(P_{\boldsymbol{\omega}}(x^{(l)}|\boldsymbol{\theta}^{(k)}))\biggr)\biggr] \\
  & \times (E^{(k)}-\overline{E^{(k)}}) \, .
\end{aligned}
\end{equation}
Sampling $\boldsymbol{\theta}^{(k)}$ from a conditional autoregressive NN is conceptually and practically appealing. Yet, the test results visualized in Fig.~\ref{vm_plot} indicate that this approach does not perform better than the gradient-based optimization described in Appendix~\ref{app2}. The accuracy shows a similar improvement rate as a function of $S$, with an approximately constant upward shift, denoting a marginally worse performance.

\section{Problem of sampling unique configurations}\label{app6}
\begin{figure}
\centering
\includegraphics[width=\columnwidth]{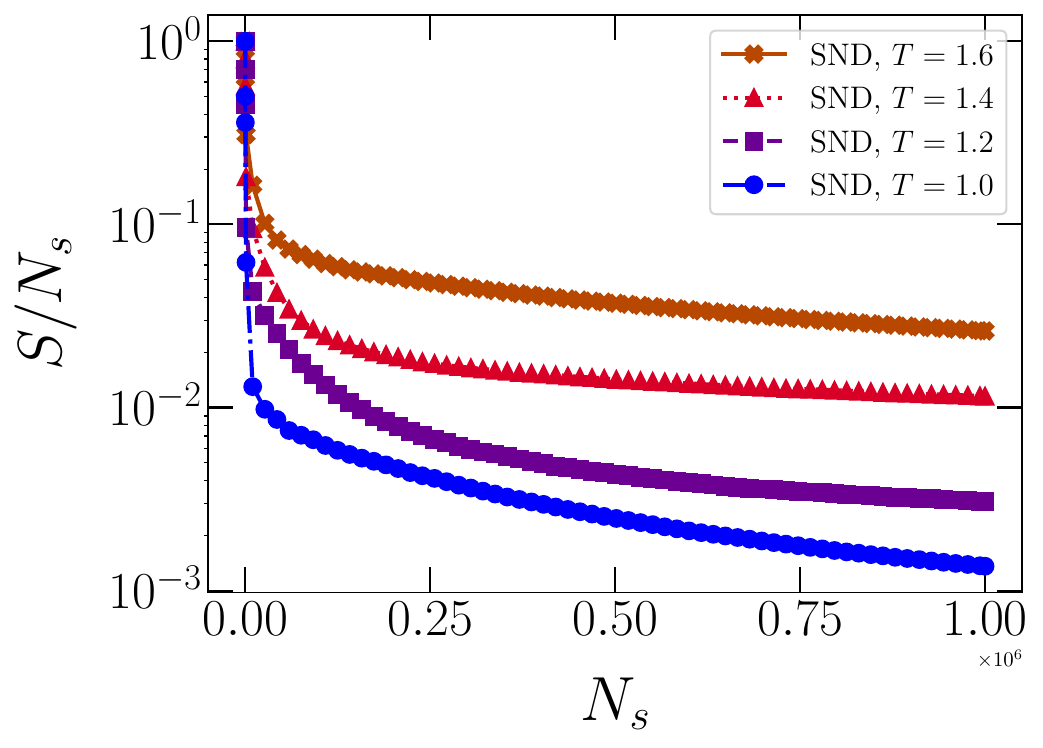}
\caption{Ratio between number of unique configurations $S$ and total number of samples $N_s$ as a function of $N_s$ for different values of the effective temperature $T$. The samples are obtained via SND for the 1D-TFIM with $N=50$ and $h=0.5$.}
\label{Ns}
\end{figure}

As the size $S$ of the configuration set increases, the probability of sampling already included configurations rapidly rises. This leads to a problematic computational cost for sampling unique configurations. To overcome this problem, we introduce the effective temperature parameter $T$~\cite{xuan2025exploringimpacttemperaturescaling}. This controls the shape of the output distribution by tuning the width of the softmax activation function in the final NN layer. The NN produces two outputs for each spin, and the softmax turns these outputs into probabilities of sampling 0 or 1. Specifically, if $Y_0$ and $Y_1$ are the two outputs, then
\begin{equation}
    \mathrm{Softmax}(Y_q) = \frac{\exp{(Y_q/T)}}{\sum_{q\in\{0,1\}}\exp{(Y_q/T)}} \, .
\end{equation}
Complementary strategies were introduced in Ref.~\cite{reinholdt2025fundamentallimitationssamplebasedquantum}.
During training, we set $T=1$, but in the inference phase we increase $T$ to have a broader distribution, thus favoring the sampling of different outputs. This effect is demonstrated in Fig.~\ref{Ns}. Indeed, while with $T=1$ the ratio between the number of unique configurations $S$ and of total configurations $N_s$ decreases almost exponentially fast, slightly larger effective temperatures suffice to significantly increase the number of unique configurations, thus drastically decreasing the computational cost of sampling. 

\begin{figure}
\centering
\includegraphics[width=\columnwidth]{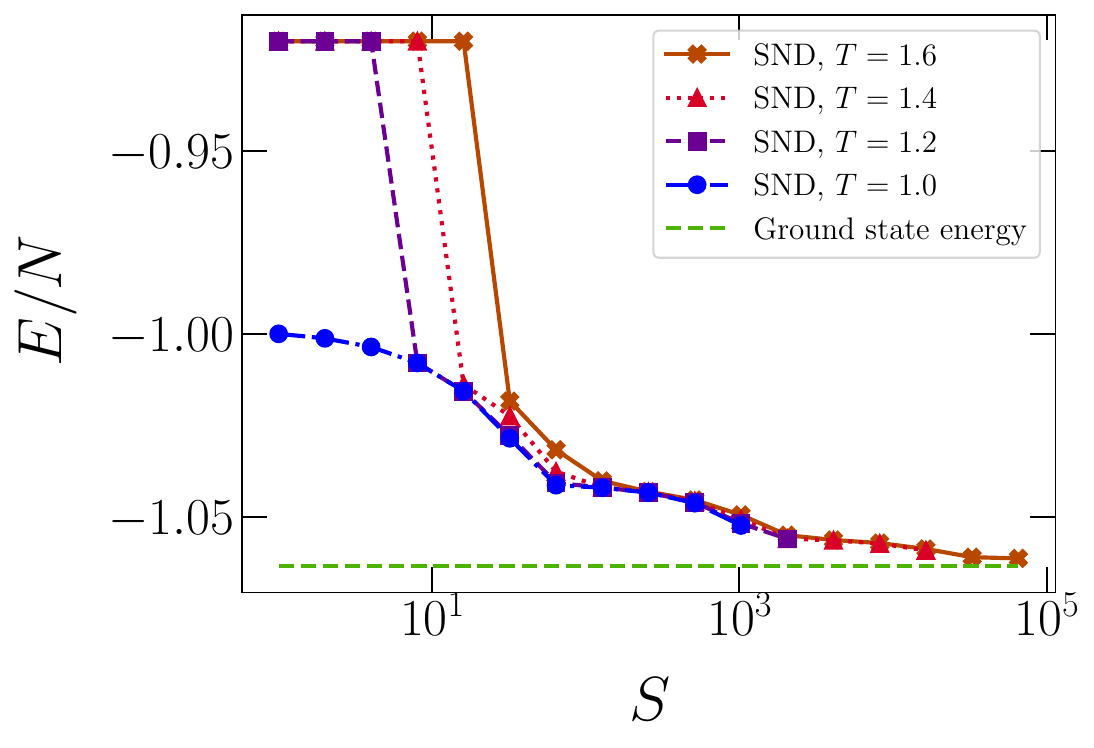}
\caption{Energy per spin $E/N$ obtained via SND as a function of the number of unique configurations $S$ used to build the subspace Hamiltonian. The different datasets correspond to different effective temperatures $T$. The testbed model is the 1D-TFIM with $N=50$ spin and transverse field $h=0.5$.}
\label{temperature}
\end{figure}

Importantly, increasing $T$ to values appropriate for efficient sampling does not reduce the performance of SND approaches. This is demonstrated in Fig.~\ref{temperature}, where one observes that, for $S \gtrsim 10^2$, values of $T\in[1,1.6]$ provide comparable accuracies for the 1D-TFIM.
\begin{figure*}
	\centering
	\includegraphics[width=\textwidth]{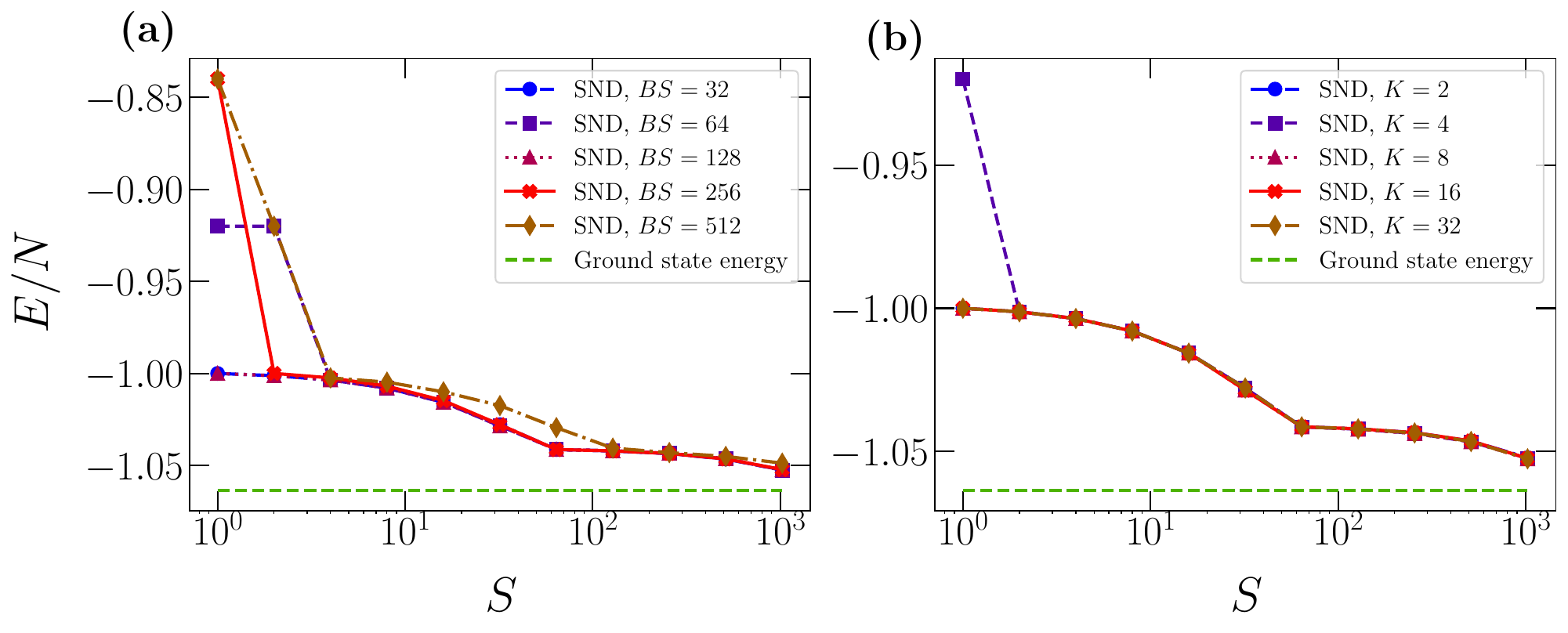}
	\caption{Energy per spin $E/N$ obtained with SND as a function of the number of unique configurations $S$ used to build the subspace Hamiltonian for the 1D-TFIM featuring $N=50$ spins and $h=0.5$. Panels (a) and (b) display results for different values of $BS$ and $K$, respectively.} 
	\label{bs_nb}
\end{figure*}
\section{Autoregressive neural network and hyperparameters}\label{app7}
We use a transformer encoder~\cite{vaswani2023attentionneed} with causal mask, two layers, four attention heads, and embedding size equal to 64. During training, we use $K=16$ batches for SND and $K=4$ batches for AB-SND, and number of sampled bitstrings equal to $BS=128$. It is worth emphasizing that the latter does not coincide with the number of unique configurations $S$ used during inference. 
%
%During training, it is not relevant whether the sampled configurations are unique. 
%
Also, the rare repeated configurations are simply discarded.
A comparison of the performances obtained with different values of $K$ and $BS$ is shown in Fig.~\ref{bs_nb}. Notably, the accuracy of the SND method is not significantly affected by different choices for these parameters.

\end{document}